\documentclass[10pt,twocolumn]{article}

\usepackage[dvips]{graphicx}
\usepackage{psfrag}
\usepackage{cite}

\usepackage[english]{babel}

\usepackage{amsmath}

\newlength{\plotwidth}
\allowdisplaybreaks[2]

\newcommand{\gev}{\operatorname{GeV}}
\newcommand{\mev}{\operatorname{MeV}}

\newcommand{\ms}{\mskip 1.5mu}

\newcommand{\jpsi}{J\mskip -2mu/\mskip -0.5mu\Psi}

\begin{document}

\title{%
\raggedleft{\normalsize DESY 07-195} \\[0.5em]
\centering\textbf{\LARGE Some numerical studies of the evolution \\
  of generalized parton distributions}}

\author{M.~Diehl and W.~Kugler \\
  \textit{Deutsches Elektronen-Synchroton DESY, 22603 Hamburg,
    Germany}}

\date{\parbox{0.9\textwidth}{\small%
  \textbf{Abstract:} We study the evolution behavior of generalized
  parton distributions at small longitudinal momentum fraction.
  Particular attention is paid to the ratio of a generalized
  parton distribution and its forward limit, to the mixing between
  quarks and gluons, and to the dependence on the squared momentum
  transfer~$t$.
}}

\maketitle


\section{Introduction}
\label{sec:intro}

A characteristic property of generalized parton distributions (GPDs)
is their renormalization scale dependence, described by evolution
equations whose derivation led to the very discovery of these
functions more than a decade ago \cite{Muller:1994fv}.  On a practical
level, the scale dependence of GPDs is of direct importance for the
quantitative description of exclusive scattering processes.  Moreover,
understanding general features of the evolution behavior should be
helpful for developing realistic models and parameterizations of GPDs.
The question how a given input distribution changes when evolved to
higher scales has been addressed in several studies, both numerically
and analytically \cite{Frankfurt:1998ha,Martin:1998wy,Shuvaev:1999ce,%
  Golec-Biernat:1999ib,Blumlein:1999sc,Musatov:1999xp,Freund:2001bf}.
Further progress has been achieved recently
\cite{Mueller:2005ed,Kirch:2005tt} by constructing explicit solutions
of the evolution equations with methods that generalize the familiar
Mellin moment inversion for parton density functions (PDFs).

The aim of the present contribution is to study a number of aspects in
the evolution of GPDs at a numerical level.  We will largely
concentrate on the value of the GPDs at $x=\xi$, which at leading
order in $\alpha_s$ determines the imaginary part of scattering
amplitudes, and via dispersion relations also gives their real part up
to a $\xi$ independent constant \cite{Anikin:2007yh}.  Furthermore we
will focus on the region of small $\xi$, where the behavior of
distributions can be conveniently approximated by a power-law
behavior.  We will pay special attention to the mixing between the
gluon GPD $H^g(x,\xi,t)$ and
\begin{equation}
  \label{def-HS}
H^{S}(x,\xi,t) = \sum_{q}^{n_f}\, \bigl[
  H^q(x,\xi,t) - H^q(-x,\xi,t) \bigr] \,,
\end{equation}
whose forward limit
\begin{equation}
H^{S}(x,0,0) = S(x) = \sum_{q}^{n_f}\, \bigl[ q(x) + \bar{q}(x) \bigr]
\end{equation}
is the familiar singlet combination of quark and antiquark PDFs.  For
comparison we will also consider $H^{u-d}(x,\xi,t) = H^u(x,\xi,t) -
H^d(x,\xi,t)$ as a representative of the non-singlet sector.

After specifying in Sect.~\ref{sec:gpd-model} the GPD model used as
initial condition for the evolution, we devote most of
Sect.~\ref{sec:t-fixed} to a quantitative study of the old question
how the ratio of GPDs and PDFs behaves when evolved to higher scales.
We shall in addition take a look at the behavior of the GPDs around
$x=\xi$.  In Sects.~\ref{sec:t-ans} and \ref{sec:t-dep} we turn to the
dependence of GPDs on the squared momentum transfer $t$.  Both
theoretical considerations \cite{Burkardt:2004bv} and lattice QCD
calculations \cite{Gockeler:2006ui} indicate that this dependence is
correlated with the one on the longitudinal variables $x$ and $\xi$.
Since evolution affects the $x$ dependence at given $\xi$ and $t$, it
also affects the $t$ dependence at given $x$ and $\xi$ in a nontrivial
fashion, which we will quantify in two model scenarios.

For our calculations we have used the numerical code of
\cite{Vinnikov:2006xw}, which provides a numerically fast and stable
implementation of GPD evolution at leading order (LO) in $\alpha_s$.
The effects of next-to-leading (NLO) and next-to-next-to-leading order
(NNLO) terms in the evolution kernels have been studied
\cite{Freund:2001bf,Mueller:2005nz} and are known to be important,
especially at small $\xi$ in the gluon and singlet sector.  This
should be kept in mind as a caveat when interpreting our results, but
we think that a study at LO is still of some relevance.  On one hand,
the arguments in \cite{Golec-Biernat:1999ib,Musatov:1999xp} about the
pattern of evolution to higher scales are based on the LO kernels, so
that this order is adequate to test the numerical validity of these
arguments.  On the other hand, evolution effects on the $t$ dependence
are barely known at all, and LO results should at least provide a
valid starting point for further investigation.

In the evolution kernels and the running coupling we take $n_f=4$ for
$m_c \le \mu < m_b$ and $n_f=5$ for $\mu \ge m_b$, with the charm
and bottom quark masses $m_c =1.3\gev$ and $m_b =4.5\gev$ used in the
CTEQ6 parton analysis \cite{Pumplin:2002vw}, which we use for
calculating the model GPDs at the starting scale of evolution.  We
furthermore follow the CTEQ6 analysis in taking the two-loop running
coupling with $\Lambda^{\smash{(4)}} = 326 \mev$ and
$\Lambda^{\smash{(5)}} = 226 \mev$, which corresponds to
$\alpha_s^{\smash{(4)}}(m_c) = 0.40$ and $\alpha_s^{\smash{(5)}}(M_Z)
= 0.118$.  We shall not consider scales below $\mu= 1.3\gev$, which we
regard as a compromise between starting evolution at a ``low scale''
and staying in a region where $\alpha_s$ is not so large that the LO
approximation becomes more and more questionable.


\section{Initial conditions}
\label{sec:gpd-model}

At the starting scale of evolution, we use the Musatov-Radyushkin
ansatz, which is based on double distributions \cite{Musatov:1999xp}.
With the conventional definitions of $H^q$ and $H^g$, given e.g.\ in
\cite{Diehl:2003ny}, we can write this ansatz as
\begin{align}
  \label{dd-models}
H^{q}(x,\xi,t) &=
\int_{-1}^1 d\beta \int_{-1+|\beta|}^{1-|\beta|} d\alpha\;
  \delta(x-\beta-\xi\alpha)\,
\nonumber \\
\phantom{\int} & \quad\times
  h_b(\beta,\alpha)\, H^{q}(\beta,0,t) \,,
\nonumber \\
H^g(x,\xi,t) &=
\int_{-1}^1 d\beta \int_{-1+|\beta|}^{1-|\beta|} d\alpha\;
  \delta(x-\beta-\xi\alpha)\,
\nonumber \\
\phantom{\int} & \quad\times
  h_b(\beta,\alpha)\,  H^g(\beta,0,t)
\end{align}
with
\begin{align}
  \label{profile}
h_b(\beta,\alpha) &= \frac{\Gamma(2b+2)}{2^{2b+1}\Gamma^2(b+1)}\,
\frac{[ (1-|\beta|)^2- \alpha^2 ]^b}{(1-|\beta|)^{2b+1}} \,.
\end{align}
In this work we will use different values of the profile parameter
$b$, which for simplicity will always be taken equal for all quark and
gluon distributions.  The ansatz \eqref{dd-models} has been
extensively used in the literature so far.  One should keep in mind
that it does not exhaust the possibilities of modeling, and other
approaches
\cite{Guzey:2005ec,Ahmad:2007vw,Kumericki:2007sa,Hwang:2007tb} are
being pursued in the literature.  As we will see, this model does
however provide enough flexibility to address a number of important
questions.

The model also permits useful analytic approximations at small $\xi$.
At $x=\xi$ the integrals in \eqref{dd-models} are restricted to $\beta
< 2\xi$, so that for $\xi\ll 1$ one can neglect the $\beta$ dependence
in $h_b(\beta,\alpha)$.  Approximating $1+\xi$ by $1$ in the
integration limits, one then has \cite{Musatov:1999xp}
\begin{align}
  \label{small-xi-approx}
& H^i(\xi,\xi,t) 
\nonumber \\
& \quad \approx \frac{1}{\xi} \int_0^{2\xi} d\beta\;
  h_b\biggl(0, 1-\frac{\beta}{\xi}\biggr)\, H^i(\beta,0,t)
\nonumber \\[0.2em]
& \quad = \frac{\Gamma(2b+2)}{\Gamma^2(b+1)}\,
  \int_0^{1} dz\, (1-z)^{b}\ms z^{b}\ms H^i(2\xi z,0,t)
\end{align}
with $i=q,g$.  We will use this approximation shortly.


\section{Evolution at fixed $t$}
\label{sec:t-fixed}

In this section we study the evolution of GPDs at a fixed value of
$t$.  We take $t=0$ and do not display this variable for brevity.  To
quantify the difference between generalized and usual parton
distributions we use the conventional skewness ratios
\begin{align}
  \label{def-ratio}
R^g(\xi,\mu) &= \frac{H^g(\xi,\xi;\mu)}{H^g(2\xi,0;\mu)} \,,
\nonumber \\
R^q(\xi,\mu) &= \frac{H^q(\xi,\xi;\mu)}{H^q(2\xi,0;\mu)} \,,
\end{align}
where we have explicitly displayed the dependence on the scale $\mu$
in the distributions.

As is well known, the PDFs obtained from fits to data follow an
approximate power-law behavior at small $x$,
\begin{align}
  \label{PDF-pow}
x g(x) &\approx a x^{-\lambda} \,,
&
x q(x) &\approx a x^{-\lambda}
\end{align}
at given $\mu$, where $a$ and $\lambda$ depend of course on the parton
species.  With the ansatz \eqref{dd-models} for GPDs this leads to a
power-law behavior
\begin{align}
  \label{GPD-pow}
H^g(\xi,\xi) &\sim \xi^{-\lambda} \,,
&
\xi H^q(\xi,\xi) &\sim \xi^{-\lambda}
\end{align}
of the GPDs at small $\xi$ according to \eqref{small-xi-approx}, with
the same powers $\lambda$ as for the corresponding PDFs.  The skewness
ratios at small $\xi$ are readily obtained as
\begin{align}
  \label{dd-ratio}
R_b^g(\lambda) &= \frac{\Gamma(2b+2)}{\Gamma(2b+2-\lambda)} \,
       \frac{\Gamma(b+1-\lambda)}{\Gamma(b+1)} \,,
\nonumber \\
R_b^q(\lambda) &= \frac{\Gamma(2b+2)}{\Gamma(2b+1-\lambda)} \,
       \frac{\Gamma(b-\lambda)}{\Gamma(b+1)}
\end{align}
at the scale $\mu$ where the ansatz \eqref{dd-models} is made.
Numerically, we find that the approximate power-laws \eqref{PDF-pow}
and \eqref{GPD-pow} remain valid under evolution to higher scales,
with powers $\lambda$ that depend on $\mu$ but remain the same for the
forward distributions and the GPDs.

Based on the considerations using the Shuvaev transformation, it has
been argued in \cite{Golec-Biernat:1999ib} that at small $\xi$ and
high enough scale, the skewness ratio should be given by
\cite{Shuvaev:1999ce,Noritzsch:2000pr}
\begin{align}
  \label{shu-ratio}
R^g_{\text{Sh}}(\lambda) &= \frac{2^{2\lambda+3}}{\sqrt{\pi}}\,
       \frac{\Gamma(\lambda+\frac{5}{2})}{\Gamma(\lambda+4)} \,,
\nonumber \\
R^q_{\text{Sh}}(\lambda) &= \frac{2^{2\lambda+3}}{\sqrt{\pi}}\,
       \frac{\Gamma(\lambda+\frac{5}{2})}{\Gamma(\lambda+3)}
\end{align}
for gluons and quarks, respectively.  Here $\lambda$ is the power in
\eqref{PDF-pow} at the scale where $R^g(\xi,\mu)$ or $R^q(\xi,\mu)$ is
evaluated.  More precisely, the ratios in \eqref{shu-ratio} are
obtained if \eqref{PDF-pow} holds and if all Gegenbauer moments of the
GPD in question are independent of $\xi$.
Musatov and Radyushkin \cite{Musatov:1999xp} have shown that at small
$x$ and $\xi$ this condition is tantamount to the GPD being given by
\eqref{dd-models} with $b = \lambda+1$, for both gluon and quark
distributions.  Indeed, one can easily check that
\begin{equation}
R^i_{\text{Sh}}(\lambda) = R^i_{\lambda+1}(\lambda) .
\end{equation}
for $i=g,q$.  Using a different line of arguments, the authors of
\cite{Kirch:2005tt} also expect that \eqref{shu-ratio} should become
valid after LO evolution to high scales, provided that one takes a
particular joint limit of large $\mu$ and $1/\xi$.
The relations \eqref{shu-ratio} are often used to calculate
high-energy scattering amplitudes, so that it is important to test
under which conditions they may be assumed to hold.

We have taken the double distribution model \eqref{dd-models} with the
CTEQ6L distributions \cite{Pumplin:2002vw} at $\mu_0 = 1.3\gev$ as
input.  After LO evolution to a scale $\mu$, we have fitted effective
power laws for $g(x)$ and $H^g(\xi,\xi)$, and we have evaluated the
skewness ratio $R^g(\xi,\mu)$ from \eqref{def-ratio}.  In analogy we
have determined power laws and ratios $R^S$ and $R^{u-d}$ for the
combinations $H^S$ and $H^{u-d}$ introduced in Sect.~\ref{sec:intro}.

\begin{figure*}
\begin{center}
  \psfrag{xx}[c][c]{\footnotesize $\mu^2\, [\gev^2]$}
  \psfrag{yy}[c][b]{\footnotesize $\lambda$}
  \psfrag{bb}{\scriptsize $10^{-3} < x < 10^{-2}$}
  \psfrag{dd}{\scriptsize $10^{-4} < x < 10^{-3}$}
  \psfrag{hh}{\scriptsize $10^{-5} < x < 10^{-4}$}
\includegraphics[width=\plotwidth]{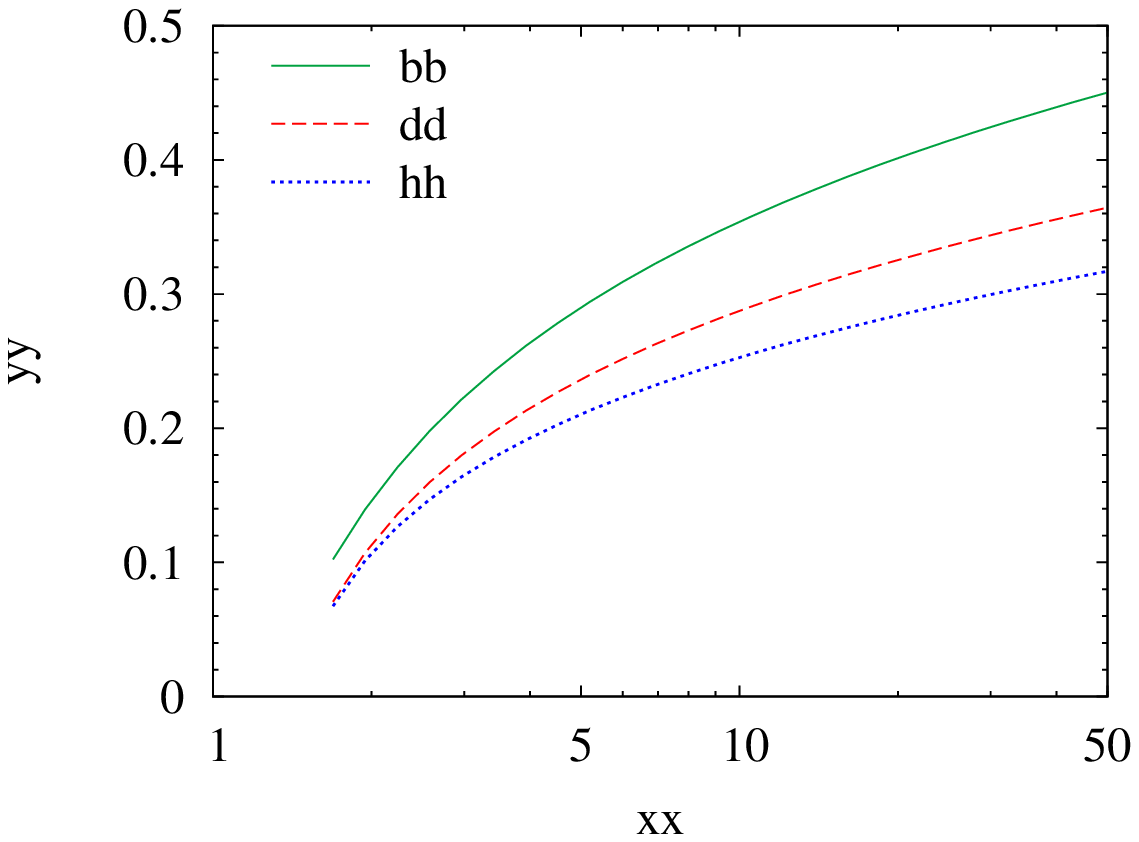}
\includegraphics[width=\plotwidth]{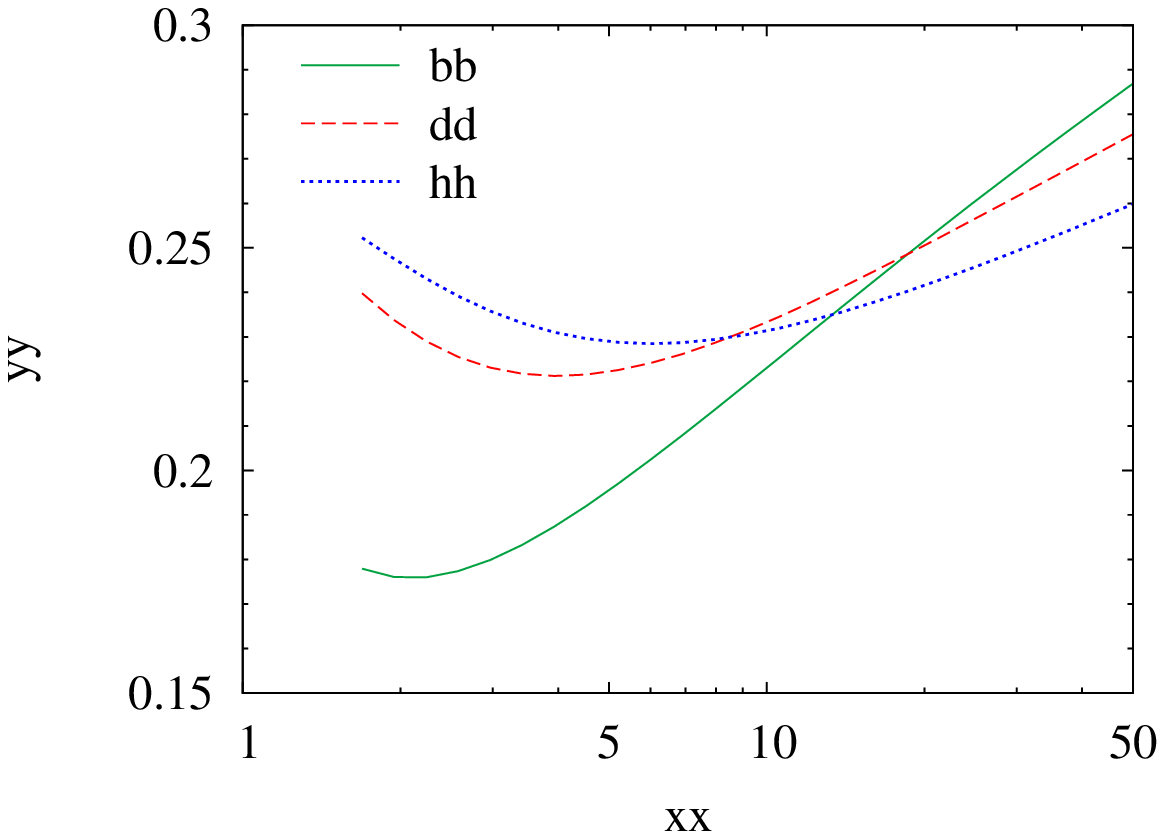}
\end{center}
\vspace{-1.5em}
\caption{\label{fig:lambda-eff1} Effective powers $\lambda$ obtained
  from fits $x g(x) \sim x^{-\lambda}$ (left) and $x S(x) \sim
  x^{-\lambda}$ (right) in three intervals of $x$ at given scale
  $\mu$.}
\end{figure*}

\begin{figure}
\begin{center}
  \psfrag{xx}[c][c]{\footnotesize $\mu^2\, [\gev^2]$}
  \psfrag{yy}[c][b]{\footnotesize $\lambda$}
  \psfrag{bb}[r][c]{\scriptsize gluon}
  \psfrag{dd}[r][c]{\scriptsize quark singlet}
\includegraphics[width=\plotwidth]{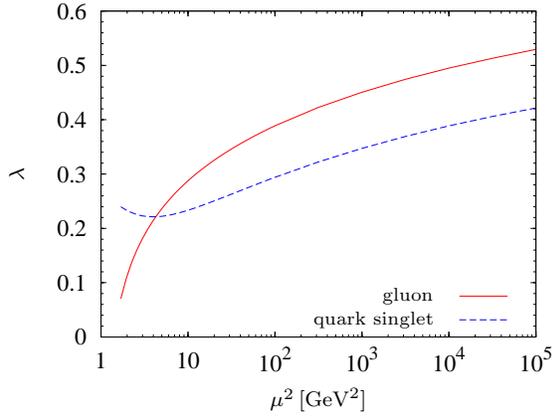}
\end{center}
\vspace{-1.5em}
\caption{\label{fig:lambda-eff2} Effective powers as in
  Fig.~\protect\ref{fig:lambda-eff1}, obtained from fits in the
  interval $10^{-4} <x< 10^{-3}$.}
\end{figure}

Let us first discuss the power-law behavior \eqref{PDF-pow} of the
PDFs, which is not exact and only valid in a certain range of $x$.  We
fitted power-laws to the CTEQ6L parameterization for $g(x)$, $S(x)$
and $u(x) - d(x)$ in the three intervals $[10^{-5}, 10^{-4}]$,
$[10^{-4}, 10^{-3}]$ and $[10^{-3}, 10^{-2}]$.  The resulting powers
for the gluon and quark singlet distributions are shown in
Fig.~\ref{fig:lambda-eff1}.  We see a clear $x$ dependence of the
effective power $\lambda$, especially at larger $x$.  In
Fig.~\ref{fig:lambda-eff2} we show the powers obtained in the
interval $10^{-4} <x <10^{-3}$ for a larger range of $\mu$.  We note
that under evolution the powers for the gluon and the quark singlet
become similar but remain different up to very high $\mu$.  This
effect has already been pointed out in \cite{Martin:1996as}.
For the non-singlet distribution $u-d$ the effective power $\lambda$
is between $-0.41$ and $-0.42$ in all three $x$ intervals.  It changes
by less than $1\%$ under evolution in the $\mu$ range corresponding to
Figs.~\ref{fig:lambda-eff1} and~\ref{fig:lambda-eff2}.

\begin{figure}
\begin{center}
  \psfrag{xx}[c][c]{\footnotesize $\xi$}
  \psfrag{yy}[c][b]{\footnotesize $H^g(\xi,\xi)$}
\includegraphics[width=\plotwidth]{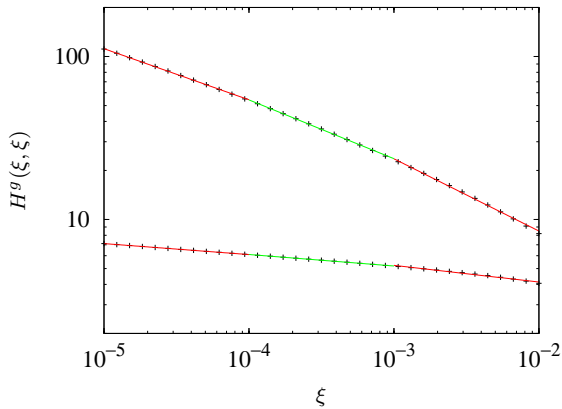}
\end{center}
\vspace{-1.5em}
\caption{\label{fig:Hxixi} Values of $H^g(\xi,\xi)$ (points) and
  power-law fits (lines) in successive intervals $[10^{-5}, 10^{-4}]$,
  $[10^{-4}, 10^{-3}]$, $[10^{-3}, 10^{-2}]$ of $\xi$.  The lower
  curve is for the starting scale $\mu_0^2=1.69 \gev^2$ and the upper
  one for $\mu^2= 50\gev^2$.}
\end{figure}

According to \eqref{small-xi-approx} there is no simple relation
between the ranges of $x$ and $\xi$ in which the same power-law
behavior should approximately hold for a PDF and the corresponding
GPD.
For simplicity we have fitted $H^g(\xi,\xi)$, $H^S(\xi,\xi)$ and
$H^{u-d}(\xi,\xi)$ to power laws \eqref{GPD-pow} in the same $\xi$
intervals that we took for the PDFs.  An example of such a fit is
shown in Fig.~\ref{fig:Hxixi}, where we see that $H^g(\xi,\xi)$ indeed
follows an approximate power-law over about one order of magnitude in
$\xi$ but not over the full range of the plot.
We find that corresponding powers $\lambda$ for PDFs and GPDs differ
by at most $3\%$ in the respective $x$ and $\mu$ ranges of
Figs.~\ref{fig:lambda-eff1} and \ref{fig:lambda-eff2}.  An exception
is the quark singlet distribution in the interval $10^{-3} <x
<10^{-2}$, where the power for the GPD is higher than that for the PDF
by $5\%$ to $10\%$.  This is not surprising, given that already in
Fig.~\ref{fig:lambda-eff1} we see a more rapid change of the effective
power at higher $x$.
For definiteness we will evaluate $R_b(\lambda)$ and
$R_{\text{Sh}}(\lambda)$ with the powers fitted to the PDFs.  We have
checked that our conclusions do not change when taking the powers for
the GPDs instead.

We note that in a specific joint limit of large $\mu$ and $1/\xi$, the
solutions of the LO evolution equations for PDFs exhibit so-called
double logarithmic scaling \cite{DeRujula:1974rf}.  In this case one
obtains $\partial \ln(x g) /\partial\ell \sim \ell^{-1/2}$ and
$\partial\ln(x S) /\partial\ell - \partial\ln(x g) /\partial\ell
\approx - (2 \ell)^{-1}$, where $\ell= \ln(x_0/x)$ with some constant
$x_0$.  The effective powers in \eqref{PDF-pow} are then larger for
the gluon than for the quark singlet distribution and depend
logarithmically on $x$.  Double logarithmic scaling for GPDs in the
region $x\ge \xi$ has been discussed in \cite{Kirch:2005tt}.

\begin{figure*}[p]
\begin{center}
  \psfrag{xx}[c][c]{\footnotesize $\mu^2\, [\gev^2]$}
  \psfrag{dd}[r][c]{\scriptsize $b=\lambda+1$}
  \psfrag{hh}[r][c]{\scriptsize $b=2$}
  \psfrag{yy}[c][b]{\footnotesize $R^g$ at $\xi=3.2\times 10^{-3}$}
\includegraphics[width=\plotwidth]{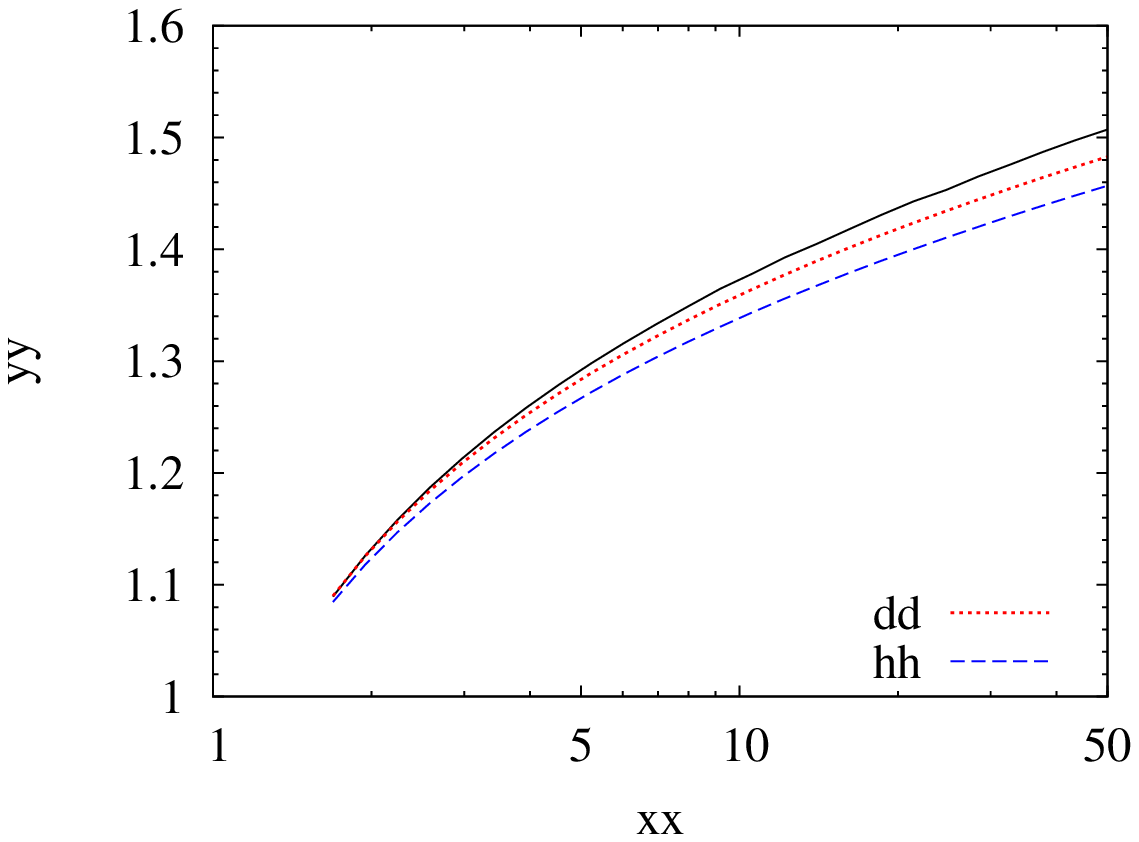}
  \psfrag{yy}[c][b]{\footnotesize $R^g$ at $\xi=3.2\times 10^{-5}$}
\includegraphics[width=\plotwidth]{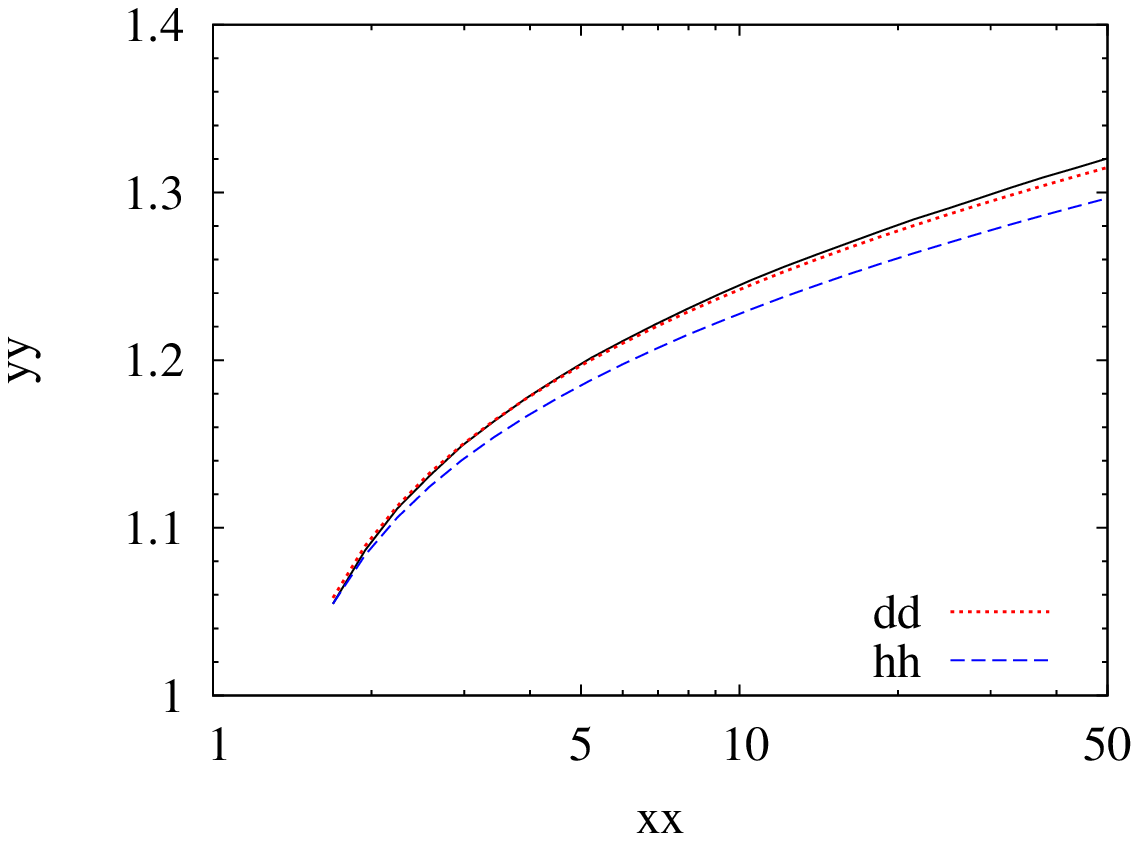}
\end{center}
\vspace{-1.5em}
\caption{\label{fig:Rg1} The skewness ratio for gluons.  Here and in
  Figs.~\protect\ref{fig:Rg2}, \ref{fig:Rq} and \ref{fig:Ru-d}, solid
  lines give the ratio $R(\xi,\mu)$ of GPDs and PDFs evolved to the
  scale $\mu$, and dashed or dotted lines give $R_b(\lambda)$
  calculated with the effective power $\lambda$ fitted at that scale.}
%
\begin{center}
  \psfrag{xx}[c][c]{\footnotesize $\mu^2\, [\gev^2]$}
  \psfrag{yy}[c][b]{\footnotesize $R^g$ at $\xi=3.2\times 10^{-4}$}
  \psfrag{bb}[r][c]{\scriptsize $b=1$}
  \psfrag{dd}[r][c]{\scriptsize $b=\lambda+1$}
  \psfrag{hh}[r][c]{\scriptsize $b=2$}
  \psfrag{kk}[r][c]{\scriptsize $b=8$}
\includegraphics[width=\plotwidth]{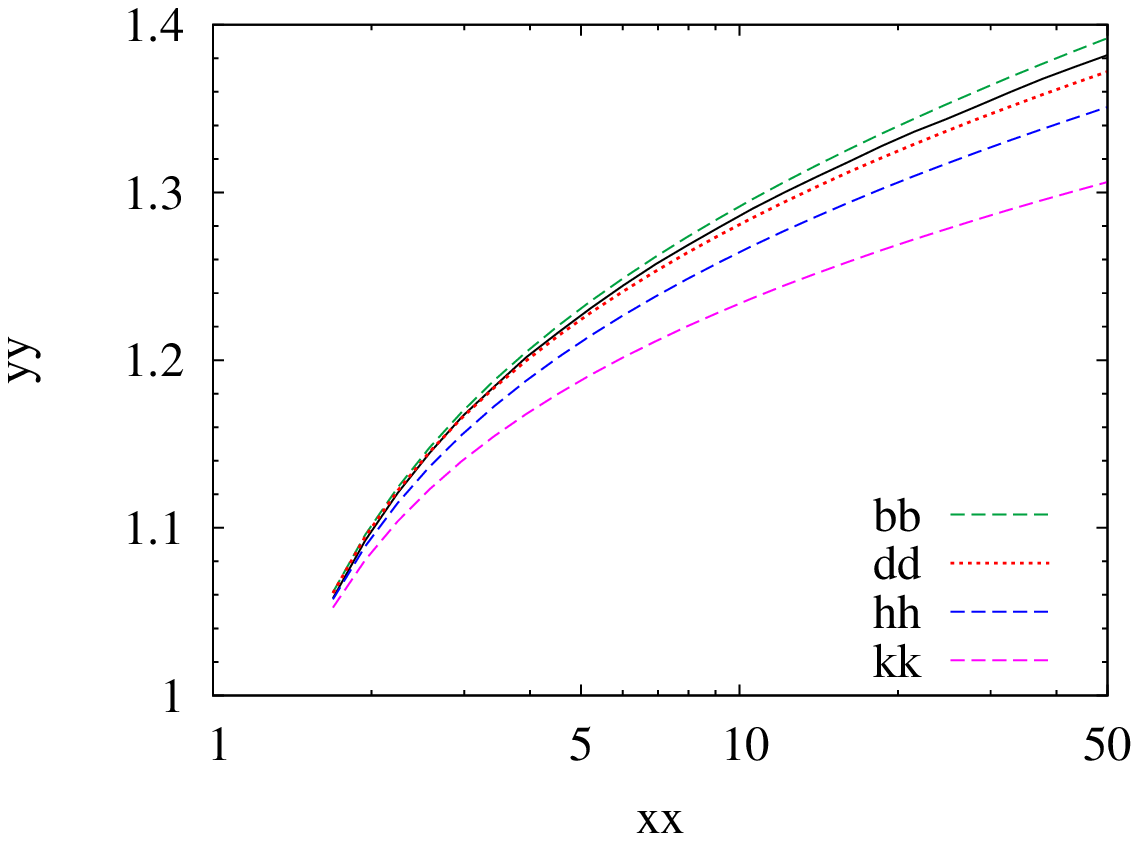}
\includegraphics[width=\plotwidth]{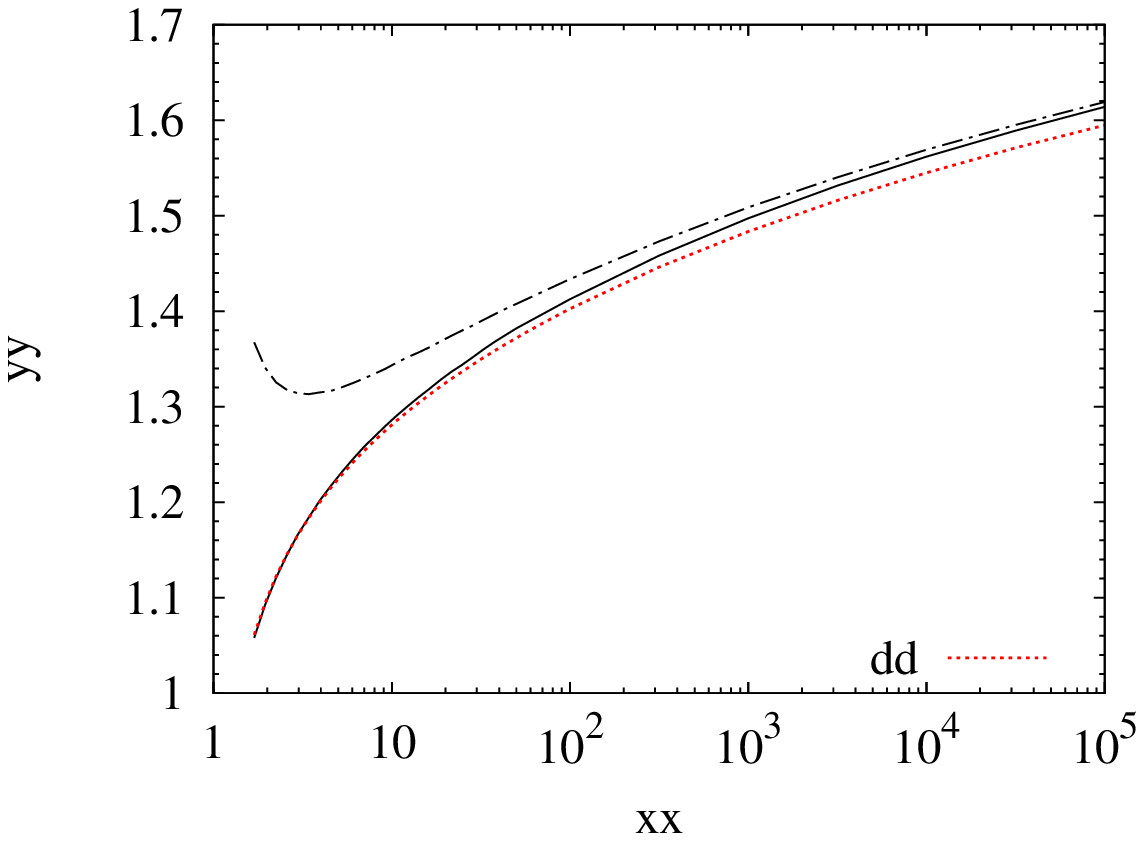}
\end{center}
\vspace{-1.5em}
\caption{\label{fig:Rg2} The skewness ratio for gluons at
  $\xi=3.2\times 10^{-4}$.  The dot-dashed curve in the right plot is
  for the alternative model \protect\eqref{alt-model} described in the
  text.}
%
\begin{center}
  \psfrag{xx}[c][c]{\footnotesize $\mu^2\, [\gev^2]$}
  \psfrag{yy}[c][b]{\footnotesize $R^S$ at $\xi=3.2\times 10^{-4}$}
  \psfrag{bb}{\scriptsize $b=1$}
  \psfrag{dd}{\scriptsize $b=\lambda+1$}
  \psfrag{hh}{\scriptsize $b=2$}
  \psfrag{kk}{\scriptsize $b=8$}
\includegraphics[width=\plotwidth]{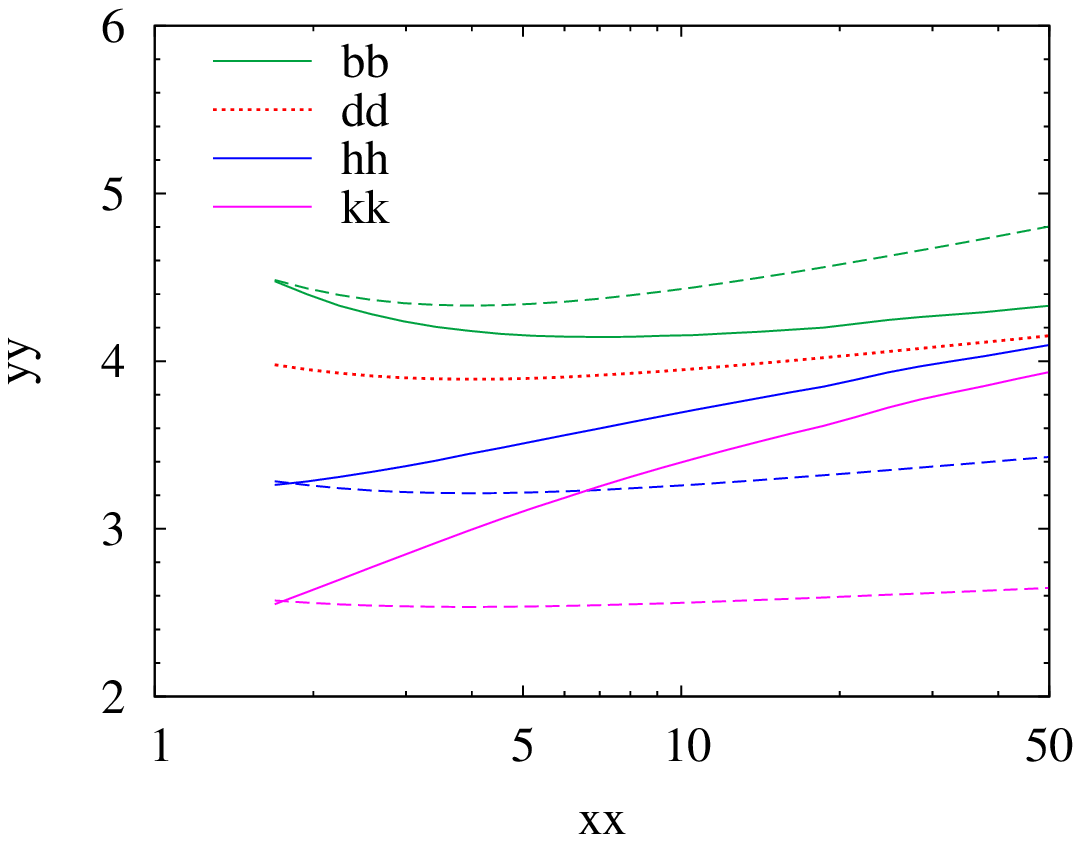}
\includegraphics[width=\plotwidth]{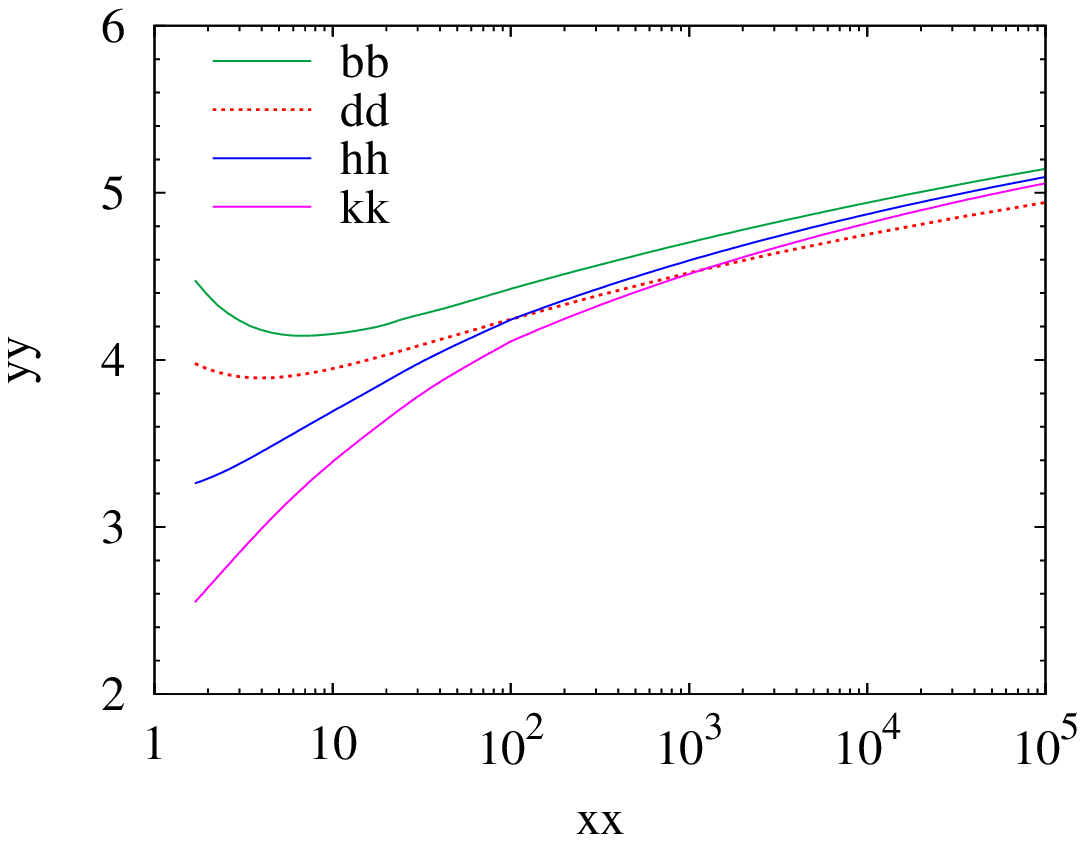}
\end{center}
\vspace{-1.5em}
\caption{\label{fig:Rq} The skewness ratio for the quark singlet
  distribution.  From top to bottom the solid curves correspond to
  $b=1, 2, 8$ in the initial condition \protect\eqref{dd-models}.  At
  the starting scale $\mu_0=1.3\gev$ they coincide with the
  corresponding dashed curves for $R^S_b(\lambda)$.}
\end{figure*}

We have evaluated the skewness ratios $R(\xi,\mu)$ from the evolved
GPDs and PDFs for $\xi = 3.2\times 10^{-5}$, $3.2\times 10^{-4}$ and
$3.2\times 10^{-3}$.  This is compared with $R_b(\lambda)$ and
$R_{\text{Sh}}(\lambda)$ calculated with $\lambda$ from our fits of
the PDFs at the corresponding scale $\mu$ and in the corresponding $x$
interval $[10^{-5}, 10^{-4}]$, $[10^{-4}, 10^{-3}]$ and $[10^{-3},
10^{-2}]$.
The result for gluons is shown in Figs.~\ref{fig:Rg1} and
\ref{fig:Rg2}, where in the initial condition we have taken $b=2$.  At
the starting scale $\mu_0 =1.3\gev$ the curves for $R^g(\xi,\mu)$ and
$R_2^g(\lambda)$ coincide as they should, whereas for increasing $\mu$
they become different.  This means that one obtains different results
for $H^g(\xi,\xi;\mu)$ when making the ansatz \eqref{dd-models} at
scale $\mu$ or when making it at scale $\mu_0$ and then evolving the
GPD.  The difference is however fairly small.

The curves for $R^g(\xi,\mu)$ and those for $R^g_{\text{Sh}}(\lambda)
= R^g_{\lambda+1}(\lambda)$ in Figs.~\ref{fig:Rg1} and \ref{fig:Rg2}
are rather close to each other.  At the starting scale they hardly
differ at all, which reflects a particularity of the model ansatz
\eqref{dd-models} for gluons.  This is because the ratio
$R^g_b(\lambda)$ in \eqref{dd-ratio} has a very weak $b$ dependence
for small $\lambda$: varying $b$ from $1$ to $\infty$ one obtains for
instance $R^g_b(0.1)$ between $1.072$ and $1.088$.  With increasing
$\lambda$ the $b$ dependence grows only slowly, with $R^g_b(0.3)$
between $1.231$ and $1.307$ for $b$ between $1$ and $\infty$.  This is
also seen in the left panel of Fig.~\ref{fig:Rg2}, where
$R^g_b(\lambda)$ is given for several $b$ values.  The solid curve in
this figure is for $b=2$ in the initial condition, but the
corresponding results for $b=1$ or $b=8$ differ by less than $0.5\%$
in the $\mu$ range of the figure.

Obviously it is hard to see whether $R^g(\xi,\mu)$ tends to
$R^g_{\text{Sh}}(\lambda)$ under evolution if the two functions are
already close at the starting scale.  To investigate this further we
take a variant of \eqref{dd-models}, namely
\begin{align}
  \label{alt-model}
H^g(x,\xi,t) &=
x \int_{-1}^1 d\beta \int_{-1+|\beta|}^{1-|\beta|} d\alpha\;
  \delta(x-\beta-\xi\alpha)\,
\nonumber \\[0.4em]
 & \quad\times
   h_b(\beta,\alpha)\,  \beta^{-1} H^g(\beta,0,t)
\end{align}
with $h_b(\beta,\alpha)$ as in \eqref{profile}.  This corresponds to a
double distribution representation for $x^{-1} H^g(x,\xi,t)$ instead
of $H^g(x,\xi,t)$, and one readily verifies that it gives Mellin
moments of $H^g(x,\xi,t)$ with a polynomial dependence on $\xi$ as
required by Lorentz invariance.  An analogous representation was first
discussed for the quark GPD of the pion \cite{Belitsky:2000vk} and was
recently found to be relevant for polarized gluon GPDs
\cite{Diehl:2007jb}.  In the case of $H^g$ the ansatz
\eqref{alt-model} has the peculiar property of giving a zero at $x=0$
that is not required by symmetry and quickly disappears under
evolution.  One may therefore not take this model too seriously, but
it serves the purpose of giving a skewness ratio sufficiently
different from the one obtained with the more conventional ansatz
\eqref{dd-models}.  This is shown in the right panel of
Fig.~\ref{fig:Rg2}, where the dot-dashed curve corresponds to initial
conditions \eqref{alt-model} for $H^g$ and \eqref{dd-models} for
$H^S$, with $b=2$ in both cases.  We see that the ratio $R^g$ in the
two models indeed tends to a common value after evolution.  This value
it not exactly equal to $R^g_{\text{Sh}}(\lambda)$ but differs from it
by less than $2\%$.  Such a small difference should not be regarded as
significant: the form \eqref{shu-ratio} of $R^g_{\text{Sh}}(\lambda)$
is obtained in \cite{Shuvaev:1999ce,Noritzsch:2000pr} from an integral
of $g(x)$ over $x$ from $\xi/2$ to $1$, assuming the power behavior
\eqref{PDF-pow} in the entire interval.  This is clearly an
approximation.

\begin{figure*}
\begin{center}
  \psfrag{xx}[c][c]{\footnotesize $x$}
  \psfrag{b}[r][c]{\scriptsize $\mu^2= 10.0 \gev^2$}
  \psfrag{d}[r][c]{\scriptsize $1.69 \gev^2$}
  \psfrag{zz}[c][b]{\footnotesize $b=1$}
\includegraphics[width=\plotwidth,%
  bb=85 50 400 300]{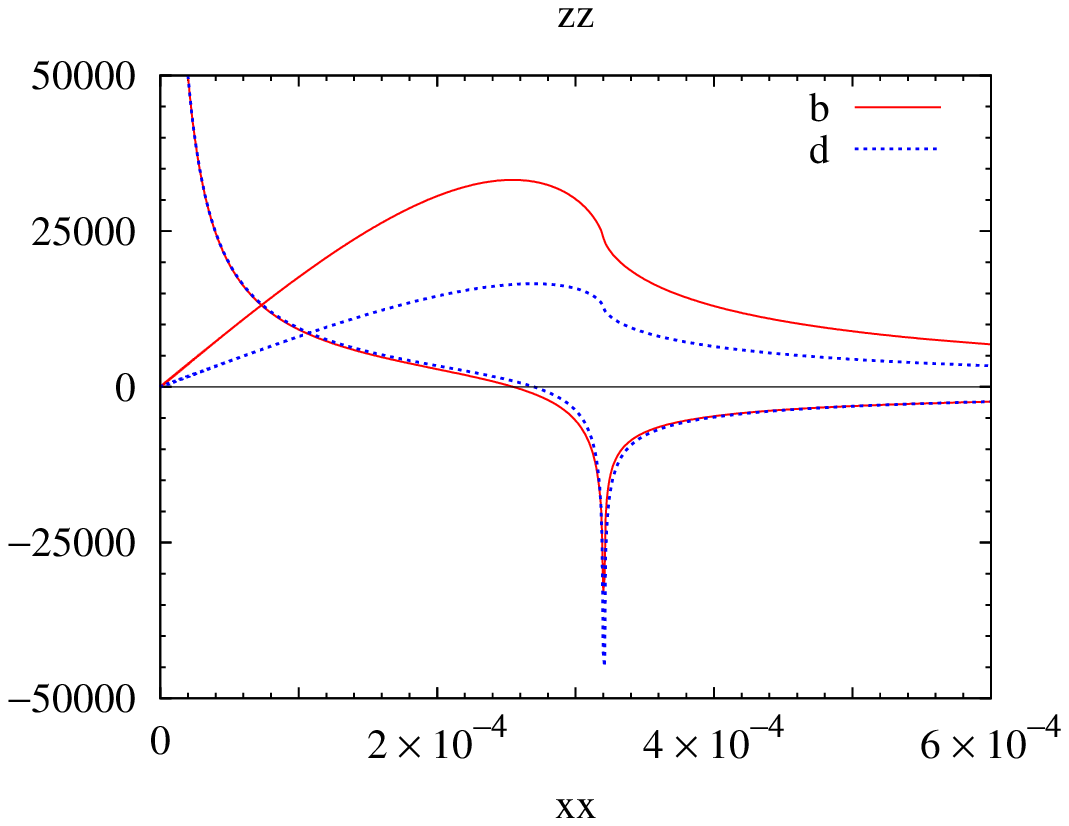}
  \psfrag{zz}[c][b]{\footnotesize $b=2$}
\includegraphics[width=\plotwidth,%
  bb=85 50 400 300]{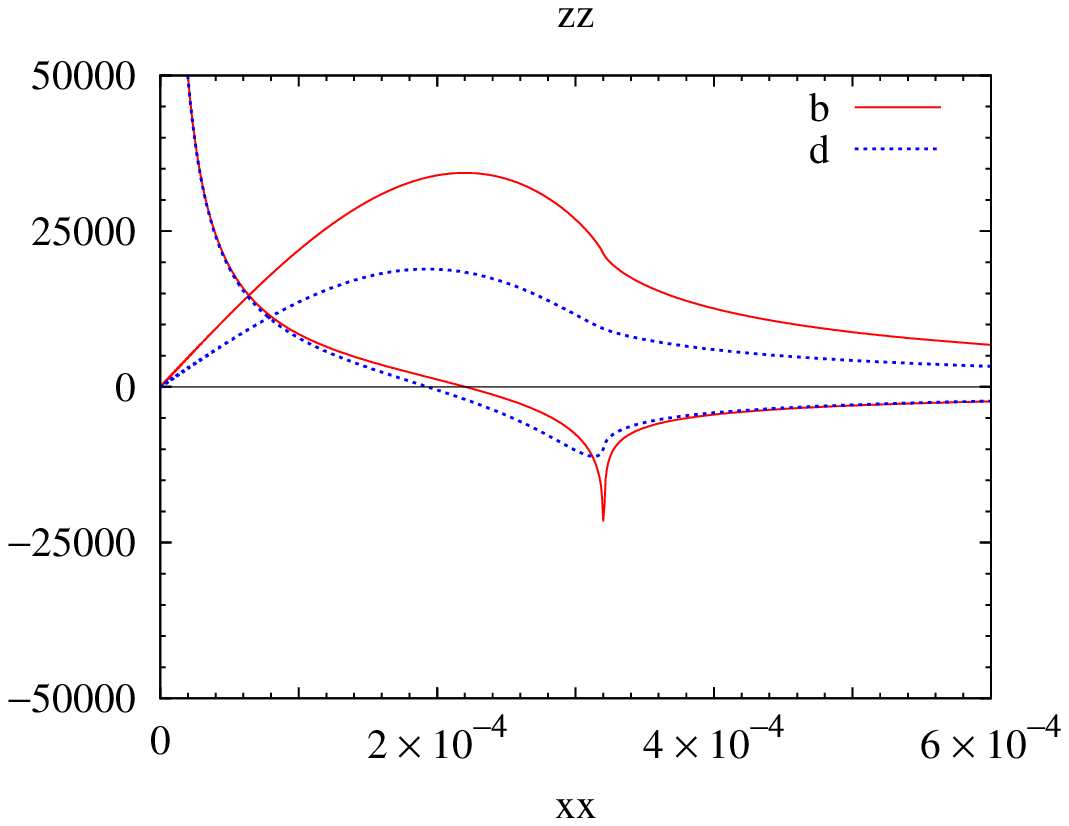}
\end{center}
\vspace{-1.5em}
\caption{\label{fig:deriv} The quark singlet GPD for $\xi=3.2\times
  10^{-4}$ at the starting scale $\mu_0^2= 1.69\gev^2$ and after
  evolution to $\mu^2= 10\gev^2$.  For each scale the upper curve
  gives $H^S(x,\xi)$ and the lower curve gives $(\partial/\partial x)
  \ln H^S(x,\xi)$.  The left plot is for $b=1$ and the right plot for
  $b=2$ in the initial condition.}
\end{figure*}

We now turn to the skewness ratio for the quark singlet distribution,
which is shown in Fig.~\ref{fig:Rq}.  In contrast to the gluon case,
different values of $b$ in the ansatz \eqref{dd-models} lead to
significantly different skewness ratios at the starting scale.
Evolution to higher $\mu$ brings the curves of $R^S(\xi,\mu)$ for
different initial conditions closer to each other.  As in the case of
gluons, they do not exactly approach the curve we calculate for
$R^S_{\text{Sh}}(\lambda)$, but again this should not be regarded as
significant since the power-law \eqref{PDF-pow} with a fixed value of
$\lambda$ is only an approximation for a certain $x$ range.

In the left panel of Fig.~\ref{fig:Rq} we also see a clear difference
between the evolved ratios $R^S(\xi,\mu)$ and the curves for
$R^S_b(\lambda)$ with $\lambda$ taken at the corresponding scale
$\mu$.  In general there is hence a notable dependence of $H^S$ on the
scale where the ansatz \eqref{dd-models} is made, especially for
larger values of $b$.  We find the dependence less pronounced for
$b=1$, in agreement with what was found in \cite{Musatov:1999xp} for
$\xi = 5.26 \times 10^{-2}$.

Based on the inversion of Gegenbauer moments, it was argued in
\cite{Kirch:2005tt} that quark distributions $H^q(x,\xi)$ should
develop a singular derivative $(\partial/\partial x) H^q(x,\xi)$ at
$x=\xi$ after evolution.  To investigate this, we have numerically
calculated $(\partial/\partial x) H^S(x,\xi)$ from the difference
quotient for successive points in $x$, which around $x=\xi$ were
spaced in intervals of $9\times 10^{-7}$.  In Fig.~\ref{fig:deriv} we
plot $H^S(x,\xi)$ together with its logarithmic derivative
$(\partial/\partial x) \ln H^S(x,\xi)$.  Taking the derivative of
\eqref{dd-models} and making the same approximations which lead to
\eqref{small-xi-approx}, one finds that $(\partial/\partial x)
H^S(x,\xi)$ is singular at $x=\xi$ for $b\le 1+\lambda$.  Indeed, we
see in the figure that at the starting scale the derivative has a
singularity for $b=1$ but remains finite for $b=2$.  Under evolution a
singularity develops for $b=2$, whereas for $b=1$ the logarithmic
derivative hardly changes.  Notice that in the curves for $H^S(x,\xi)$
one can barely recognize that the tangent at $x=\xi$ should be
vertical: this illustrates the limitations of rendering a weakly
singular derivative in a plot.
In contrast to the quark case, the ansatz \eqref{dd-models} with $b=1$
or $b=2$ gives a finite value of $(\partial/\partial x) H^g(x,\xi)$ at
$x=\xi$.  We have checked numerically that for both initial conditions
the derivative remains finite under evolution, in agreement with what
one expects from the analytical representation in \cite{Kirch:2005tt}.

\begin{figure}
\begin{center}
  \psfrag{xx}[c][c]{\footnotesize $\mu^2\, [\gev^2]$}
  \psfrag{yy}[c][b]{\footnotesize $R^{u-d}$ at $\xi=3.2\times 10^{-4}$}
  \psfrag{bb}[r][c]{\scriptsize $b=1$}
  \psfrag{dd}[r][c]{\scriptsize $b=\lambda+1$}
  \psfrag{hh}[r][c]{\scriptsize $b=2$}
  \psfrag{kk}[r][c]{\scriptsize $b=8$}
\includegraphics[width=\plotwidth]{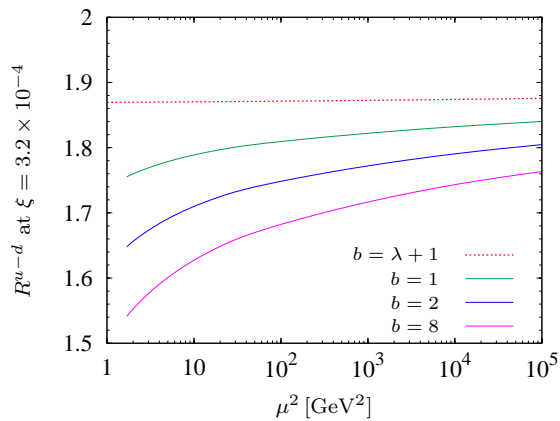}
\end{center}
\vspace{-1.5em}
\caption{\label{fig:Ru-d} The skewness ratio for the non-singlet
  distribution $H^{u-d}$.  The meaning of the curves is as in the
  right panel of Fig.~\protect\ref{fig:Rq}.}
\end{figure}

To conclude this section, we briefly investigate the quark non-singlet
distribution $H^{u-d}$.  In Fig.~\ref{fig:Ru-d} we show the skewness
ratio for different values of $b$ in the initial condition.  The
corresponding curves for $R^{u-d}_b(\lambda)$ are not shown: they
coincide with those for $R^{u-d}(\xi,\mu)$ at the starting scale and
then remain essentially flat since the effective power $\lambda$
hardly changes with $\mu$ in this case.  Under evolution to high
scales the curves for different $b$ approach each other and the one
for $R^{u-d}_{\text{Sh}}(\lambda)$, although much more slowly than for
$R^g$ or~$R^S$.

\section{Ansatz for the $t$ dependence}
\label{sec:t-ans}

To investigate the change of the $t$ dependence with evolution, we
will use the model \eqref{dd-models} with $b=2$ for quarks and gluons.
We thus need an ansatz for the GPDs at zero skewness $\xi$ but finite
$t$, which is described in this section.  In all cases we assume an
exponential $t$ dependence that is correlated with $x$.  For the
valence-type combination of GPDs we take the form proposed in
Ref.~\cite{Diehl:2004cx}:
\begin{align}
  \label{DFJK4-ansatz}
H_v^{q}(x,0,t) &= H^{q}(x,0,t) + H^{q}(-x,0,t)
\nonumber \\
 & = q_v(x)\ms \exp\bigl[ t f_q(x) \bigr]
\end{align}
with $x>0$, $q_v(x) = q(x) - \bar{q}(x)$ and
\begin{align}
  \label{DFJK4-profile}
f_{q}(x) &= \alpha'_v (1-x)^3 \ln\frac{1}{x} 
\nonumber \\
& \quad {}+ B_{q} (1-x)^3 + A_{q}\ms x (1-x)^2 \,.
\end{align}
The values
$\alpha_v' = 0.9 \gev^{-2}$, $B_{u} = B_{d} = 0.59 \gev^{-2}$, $A_{u}
= 1.22 \gev^{-2}$ and $A_{d} = 2.59 \gev^{-2}$
together with the CTEQ6M parameterization for $q_v(x)$ at $\mu= 2\gev$
lead to a good description of the data for the Dirac form factors
$F_1(t)$ of proton and neutron, which are obtained by combining $\int
dx\, H_v^q(x,0,t)$ for $u$ and $d$ quarks with the appropriate
charge factors.

For small $x$ we can approximate \eqref{DFJK4-profile} as
$f_q(x) \approx \alpha'_v \ln(1/x) + B_q$
and thus have
\begin{align}
  \label{q-approx-1}
H_v^q(x,0,t)
  &\approx q_v(x)\, x^{-t \alpha'_v}\, e^{t B_q}
\nonumber \\[0.2em]
  &\approx a\ms x^{- (1 + \lambda + t \alpha'_v)}\, e^{t B_q} \,,
\end{align}
where in the second step we have assumed a small-$x$ behavior of the
valence quark distributions as in \eqref{PDF-pow}.  Since the $x$
dependence of \eqref{q-approx-1} is a power-law, the integral in
\eqref{small-xi-approx} can be performed as in
Sect.~\ref{sec:t-fixed}, and we can use \eqref{dd-ratio} for the
skewness ratio at small $\xi$ after replacing $\lambda$ with $\lambda
+ t \alpha'_v$.  For $b=2$ this gives
\begin{align}
  \label{q-approx-2}
& H_v^q(\xi,\xi,t) \approx
  q_v(2\xi)\ms \exp\bigl[ t f_q(2\xi) \bigr]
  \phantom{\frac{1}{1}}
\nonumber \\
& \quad\times \frac{60}{(2 - \lambda - t\alpha'_v)
        (3 - \lambda - t\alpha'_v) (4 - \lambda - t\alpha'_v)} \,.
\end{align}
For small $t$ we can write
\begin{align}
  \label{frac-approx}
\frac{1}{n - \lambda - t\alpha'} &= \frac{1}{n - \lambda}
  \exp\biggl[ -\ln\Bigl( 1 - \frac{t\alpha'}{n-\lambda} \Bigr) \biggr]
\nonumber \\
& \approx \frac{1}{n - \lambda}
  \exp\biggl[\ms t\ms \frac{\alpha'}{n-\lambda} \ms\biggr]
\end{align}
and thus approximate \eqref{q-approx-2} by
\begin{equation}
  \label{t-approx-q}
H_v^q(\xi,\xi,t) \approx H_v^q(\xi,\xi,0)\,
  \exp\bigl[ t \bar{f}_q(\xi) \bigr]
\end{equation}
with
\begin{equation}
  \label{fq-bar}
\bar{f}_q(\xi) = \alpha'_v \ln\frac{1}{\xi} + \bar{B}_q \,,
\end{equation}
where
\begin{equation}
  \label{Bq-bar}
\bar{B}_q = B_q + \alpha'_v\,
  \biggl(\ms \sum_{n=2}^4 \frac{1}{n-\lambda} - \ln 2 \biggr) \,.
\end{equation}

Turning to the gluon distribution, we take
\begin{equation}
H^g(x,0,t) = x g(x)\ms \exp\bigl[ t f_g(x) \bigr]
\end{equation}
with the function
\begin{equation}
  \label{gluon-profile}
f_g(x) = \alpha'_g (1-x)^2 \ln\frac{1}{x} + B_g (1-x)^2 \,,
\end{equation}
which has one parameter less than its counterpart
\eqref{DFJK4-profile}.  Since most phenomenological information about
gluons presently comes from small-$x$ data, it would be difficult to
constrain a third parameter.  The analog of \eqref{q-approx-2} reads
\begin{align}
  \label{g-approx}
& H^{g}(\xi,\xi,t) \approx
   2\xi\ms g(2\xi)\ms \exp\bigl[ t f_g(2\xi) \bigr]
\nonumber \\
& \quad\times \frac{60}{(3 - \lambda - t\alpha'_g)
        (4 - \lambda - t\alpha'_g) (5 - \lambda - t\alpha'_g)}
\end{align}
for $b=2$ and was already used in \cite{Goloskokov:2006hr}.  With
the approximation in \eqref{frac-approx} we find
\begin{equation}
  \label{t-approx-g}
H^{g}(\xi,\xi,t) \approx H^{g}(\xi,\xi,0)\,
  \exp\bigl[ t \bar{f}_g(\xi) \bigr] \,,
\end{equation}
where
\begin{equation}
  \label{fg-bar}
\bar{f}_g(\xi) = \alpha'_g \ln\frac{1}{\xi} + \bar{B}_g
\end{equation}
and
\begin{equation}
  \label{Bg-bar}
\bar{B}_g = B_g + \alpha'_g\,
  \Biggl(\ms \sum_{n=3}^5 \frac{1}{n-\lambda} - \ln 2 \Biggr)
\end{equation}
in analogy to the quark case.  For our numerical study we take the
parameters
$\alpha'_{\smash[b]{g}} = 0.164 \gev^{-2}$ and $B_g = 1.2 \gev^{-2}$
in order to match recent H1 data on $\jpsi$ photoproduction, whose $t$
dependence is well fitted by
\begin{equation}
  \label{h1-jpsi}
\frac{d\sigma}{dt} \propto \exp\biggl[ 
  \left( b_0 + 4\alpha'_g \ln\frac{W_{\gamma p}}{W_0} \right) t 
\,\biggr]
\end{equation}
with values $b_0 = 4.63 \gev^{-2}$ and $\alpha'_g=0.164 \gev^{-2}$ for
$W_{0} = 90 \gev$ \cite{Aktas:2005xu}.
To connect \eqref{h1-jpsi} with \eqref{t-approx-g} we have used the
approximate relation $d\sigma/dt \propto |H^g(\xi,\xi,t)|^2$, which is
obtained at tree level when one keeps only the imaginary part of the
scattering amplitude.  The skewness variable is given by $2 \xi =
(M_{\jpsi} /W_{\gamma p})^2$ in terms of the $\gamma p$ c.m.\ energy.
For simplicity, we have omitted the terms with $1/(n-\lambda)$ in
\eqref{Bg-bar} when fixing $B_g$.  For typical values of $\lambda$
they are quite small.

For antiquarks we set
\begin{align}
  \label{sea-quarks}
H^{q}(-x,0,t) &= -\bar{q}(x) \exp\bigl[ t f_{\bar{q}}(x) \bigr]
\end{align}
with $x>0$.  Little is known to date about the $t$ dependence in the
sea quark sector.  Constraints can be provided by deeply virtual
Compton scattering \cite{Aaron:2007cz,zeus-dvcs}, which at small $x$
is sensitive to both sea quark and gluon distributions.  A
comprehensive analysis of this data, as has recently been performed in
\cite{Kumericki:2007sa}, is beyond the scope of this work.  We will
instead explore the pattern of evolution for two extreme choices.  In
model~1 we set the $t$ slope $f_{\bar{q}}$ equal to the one for
valence quarks:
\begin{align}
  \label{model-1}
f_{\bar{u}} &= f_{u} \,, &
f_{\smash{\bar{d}}} &= f_{d} \,, &
f_{\bar{s}} &= f_{d} \,,
\end{align}
where the choice $f_{\bar{s}} = f_{d}$ has no strong motivation, but
does not strongly influence the results we will obtain.  In model 2 we
set instead
\begin{align}
  \label{model-2}
f_{\bar{q}} &= f_{g}
\end{align}
for all quark flavors.  The initial conditions for evolution of the
singlet and non-singlet combinations are then obtained from
\begin{align}
  \label{init-cond}
H^S(x,0,t) &= \sum_{q=u,d,s}
  \bigl[ H_v^{q}(x,0,t) - 2H^{q}(-x,0,t) \bigr] \,, 
\nonumber \\[0.3em]
H^{u-d}(x,0,t) &= H_v^{u}(x,0,t) - H^{u}(-x,0,t)
\nonumber \\[0.2em]
       & \quad  - H_v^{d}(x,0,t) + H^{d}(-x,0,t) \,.
\end{align}
For the evolution study in the next section, we make the ansatz
\eqref{dd-models} with the CTEQ6M parton distributions at $\mu_0
=2\gev$, so that we can use the fit of \cite{Diehl:2004cx} for the $t$
dependence of $H_v^{q}(x,0,t)$ as specified at the beginning of this
section.  In \eqref{init-cond} we have neglected the tiny charm quark
distribution at $\mu_0$.  To explore the region of lower scales, we
will also consider backward evolution.


\section{Evolution of the $t$ dependence}
\label{sec:t-dep}

\begin{figure*}
\begin{center}
  \psfrag{xx}[c][c]{\footnotesize $-t\; [\gev^2]$}
  \psfrag{k}[r][c]{\scriptsize $\mu^2= 50 \gev^2$}
  \psfrag{h}[r][c]{\scriptsize $4 \gev^2$}
  \psfrag{d}[r][c]{\scriptsize $3 \gev^2$}
  \psfrag{b}[r][c]{\scriptsize $2 \gev^2$}
  \psfrag{yy}[c][t]{\footnotesize $H^g(\xi,\xi,t)$}
\includegraphics[width=\plotwidth]{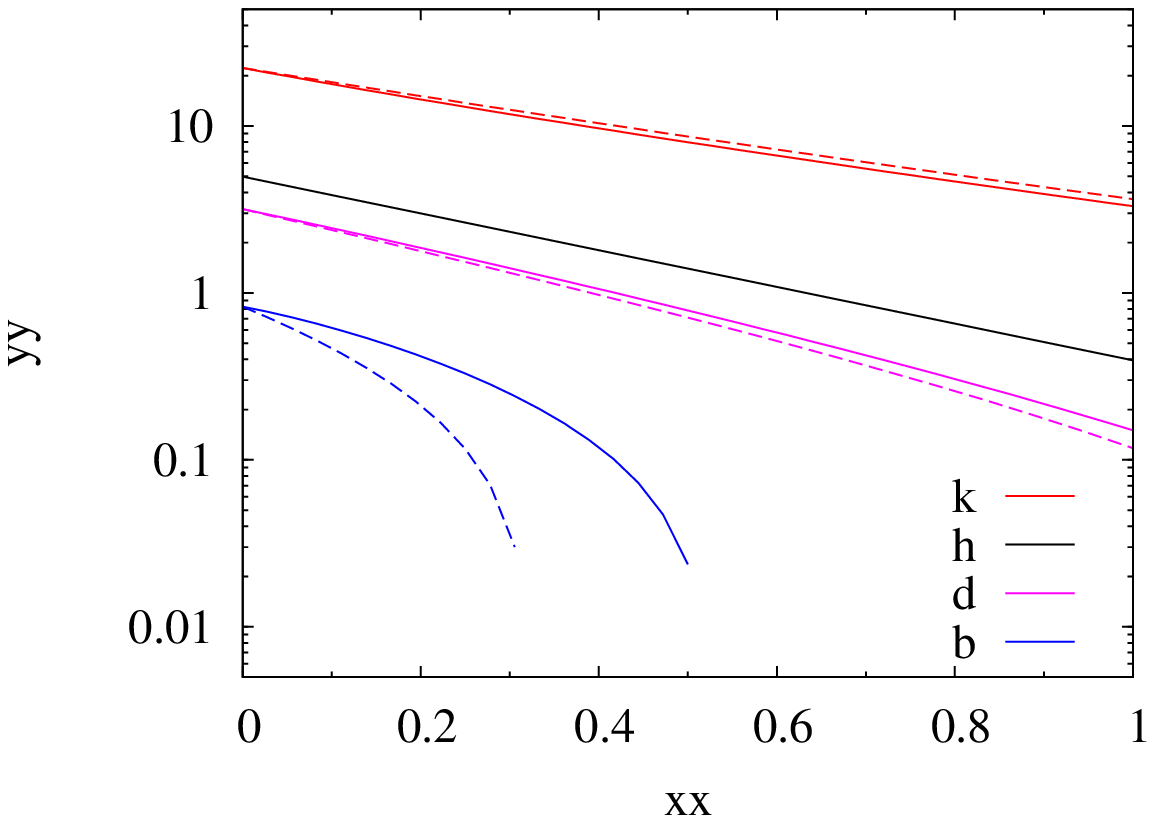}
  \psfrag{kkk}[r][c]{\scriptsize $50$}
  \psfrag{hkk}[r][c]{\scriptsize $6$}
  \psfrag{hhh}[r][c]{\scriptsize $4$}
  \psfrag{ddd}[r][c]{\scriptsize $3$}
  \psfrag{bbb}[r][c]{\scriptsize $2$}
  \psfrag{yy}[c][t]{\footnotesize $10^{-3}\times H^S(\xi,\xi,t)$}
\includegraphics[width=\plotwidth]{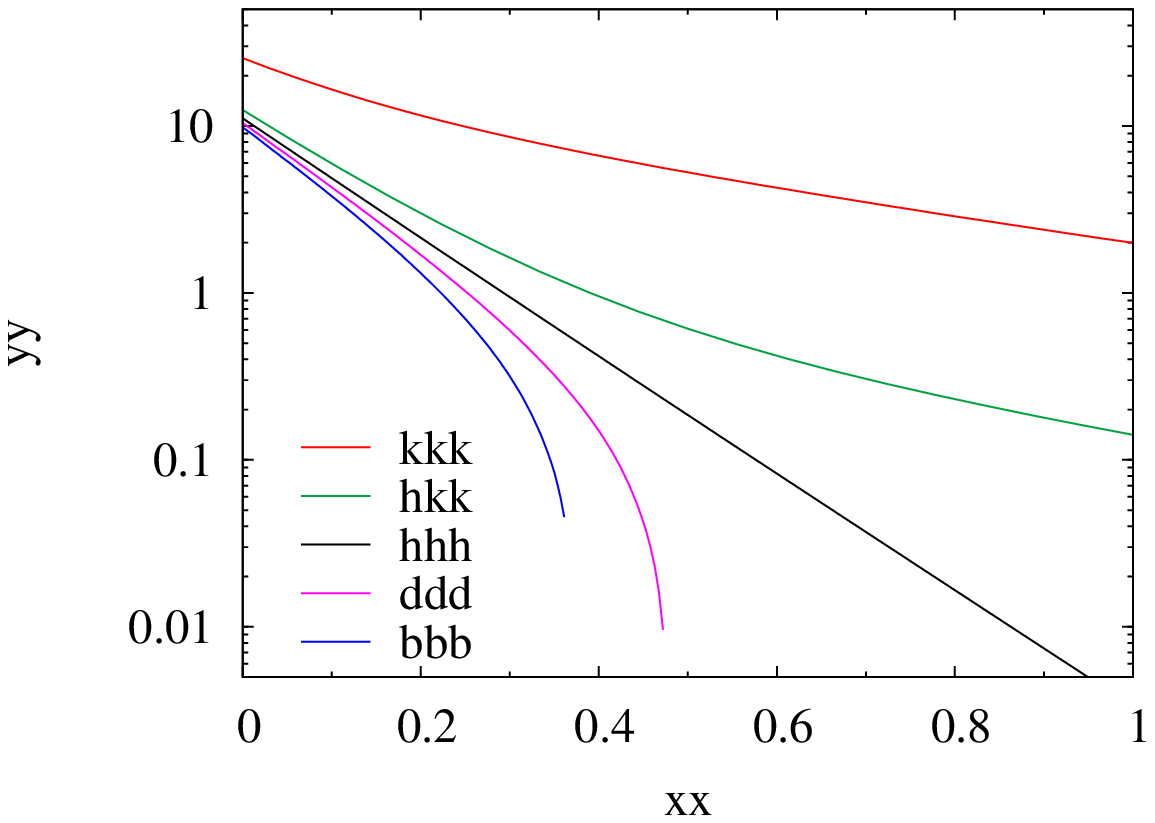}
\end{center}
\vspace{-1.5em}
\caption{\label{fig:H-t} $H^g(\xi,\xi,t)$ and $H^S(\xi,\xi,t)$ at
  $\xi= 3.2\times 10^{-4}$ for different scales $\mu^2$ as indicated.
  Here and in the following figures, solid curves are for model 1 and
  dashed curves for model 2.  At the starting scale $\mu^2 =4 \gev^2$
  the curves for $H^g(\xi,\xi,t)$ coincide in both models.}
%
\begin{center}
  \psfrag{xx}[c][c]{\footnotesize $\xi$}
  \psfrag{b}[r][c]{\scriptsize $\mu^2= 2 \gev^2$}
  \psfrag{d}[r][c]{\scriptsize $4 \gev^2$}
  \psfrag{h}[r][c]{\scriptsize $50 \gev^2$}
  \psfrag{yy}[c][t]{\footnotesize $\bar{f}_g(\xi;\mu)\, [\gev^{-2}]$}
\includegraphics[width=\plotwidth]{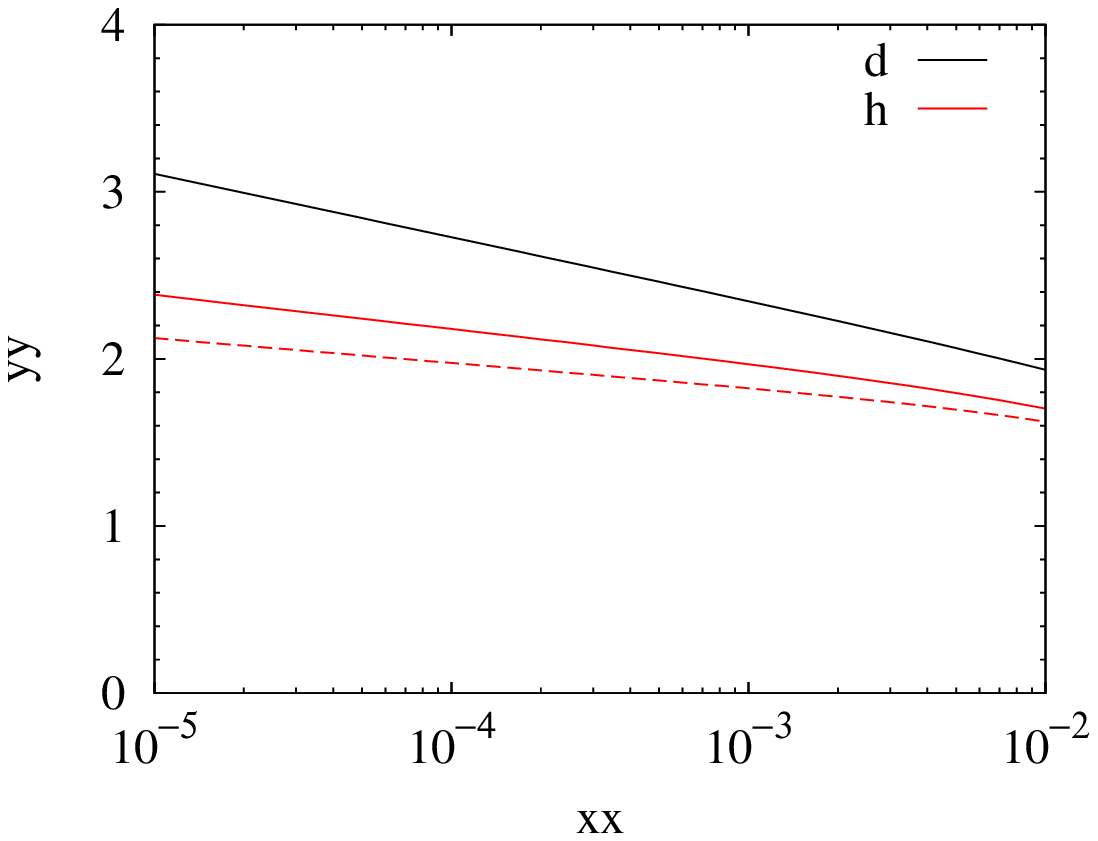}
  \psfrag{yy}[c][t]{\footnotesize $\bar{f}_S(\xi;\mu)\, [\gev^{-2}]$}
\includegraphics[width=\plotwidth]{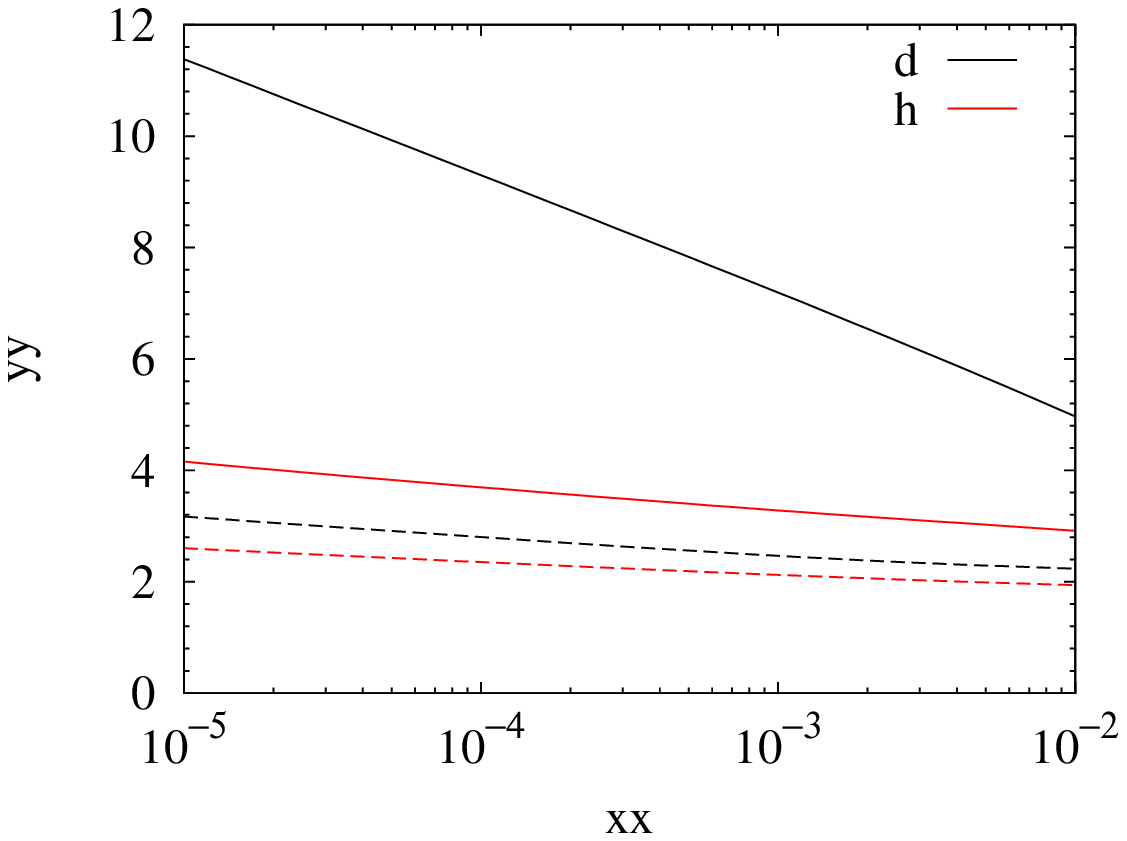}
\end{center}
\vspace{-1.5em}
\caption{\label{fig:B-vs-xi} The $t$ slope $\bar{f}(\xi;\mu)$ fitted
  according to \protect\eqref{fit-t} for the gluon (left) and the
  quark singlet GPD (right).  At $\mu^2= 4 \gev^2$ the curves for
  $\bar{f}_g(\xi;\mu)$ coincide in models 1 and 2.}
%
\begin{center}
  \psfrag{xx}[c][c]{\footnotesize $\mu^2\, [\gev^2]$}
  \psfrag{b}[r][c]{\scriptsize $i=S$}
  \psfrag{d}[r][c]{\scriptsize $g$}
  \psfrag{yy}[c][b]{\footnotesize $\bar{f}_i(\xi;\mu)\, [\gev^{-2}]$}
\includegraphics[width=\plotwidth]{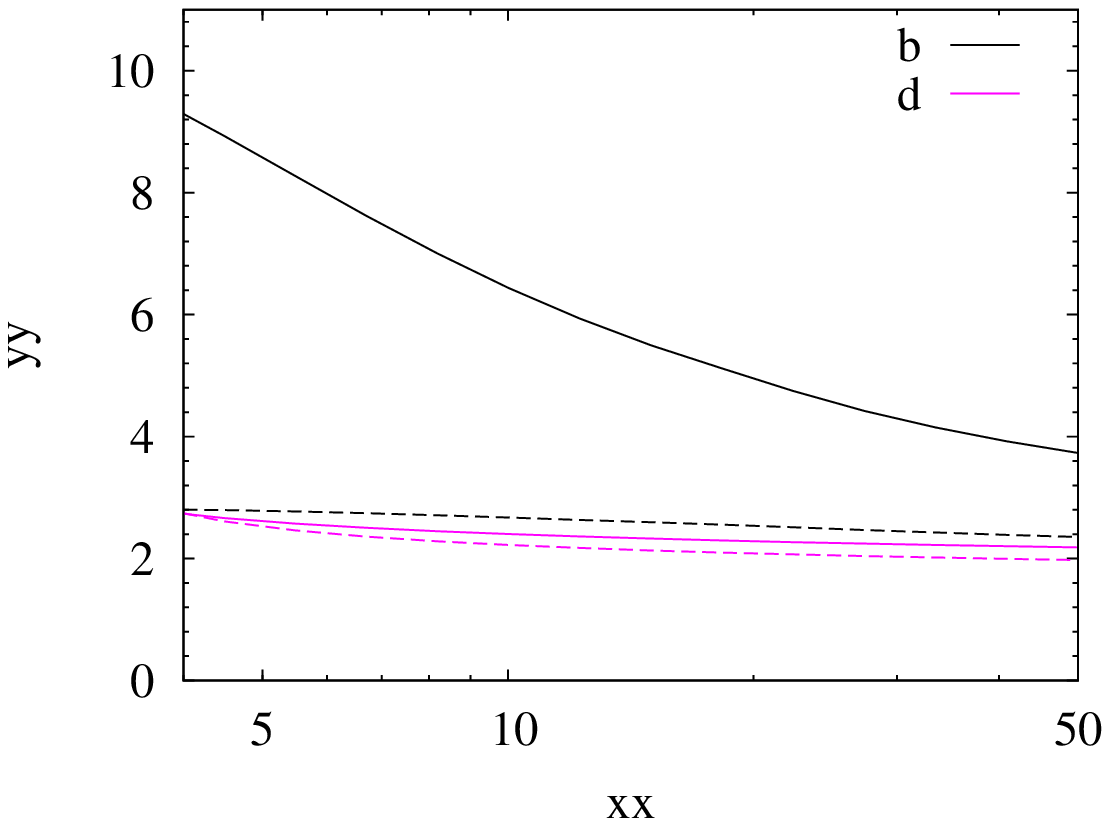}
  \psfrag{yy}[c][b]{\footnotesize $\alpha'_i(\mu)\, [\gev^{-2}]$}
\includegraphics[width=\plotwidth]{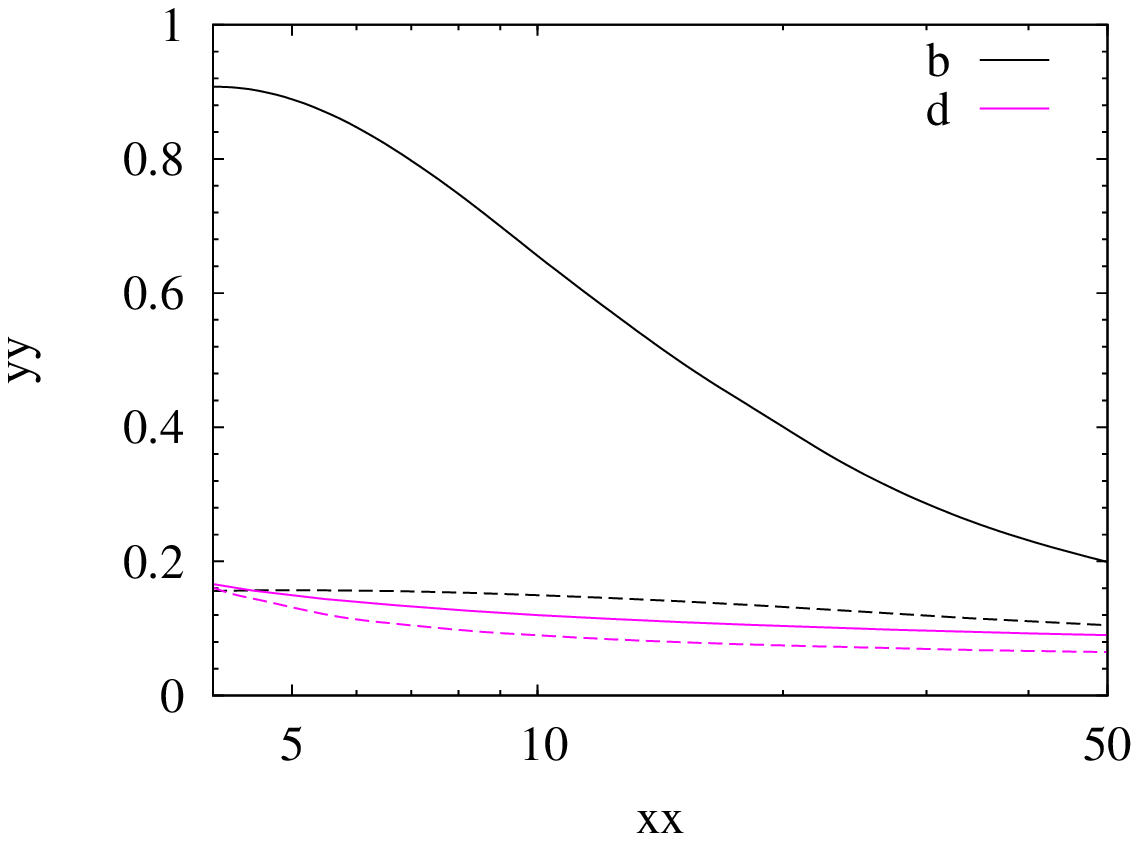}
\end{center}
\vspace{-1.5em}
\caption{\label{fig:apBxi} The $t$ slope $\bar{f}_i(\xi;\mu)$ at
  $\xi=10^{-4}$ and the shrinkage parameter $\alpha'_i$ in
  \protect\eqref{fit-xi} as a function of $\mu^2$.}
\end{figure*}

In accordance with the analytical considerations in the previous
section, we find that at the initial scale the $t$ dependence of
$H^g(\xi,\xi,t)$ is well described by an exponential form at small $t$
and $\xi$.  Evolving to higher scales we still find an approximately
exponential behavior in both model 1 and 2, as shown in
Fig.~\ref{fig:H-t} for $\xi= 3.2\times 10^{-4}$.  A slight departure
from an exact exponential in the full region $0 \le -t \le 1 \gev^2$
is however visible at $\mu^2= 50\gev^2$.  Evolving to lower scales, we
still find an approximate exponential $t$ dependence at $\mu^2=
3\gev^2$, but for yet lower scales the situation changes.  At $\mu^2=
2\gev^2$ the distribution $H^g(\xi,\xi,t)$ turns negative for $-t$
around $0.5\gev^2$ in model 1 and around $0.3\gev^2$ in model 2,
whereas at $\mu^2= 1.69\gev^2$ we have $H^g(\xi,\xi,t) < 0$ already
for $t=0$.  This is due to the behavior of the CTEQ6M gluon density at
low scales.  Since the gluon distribution in this region varies
considerably between different global parton fits, we shall not
elaborate on this issue further here.

The singlet distribution $H^S(\xi,\xi,t)$ is again well approximated
by an exponential in $t$ at the starting scale, and it stays
exponential to high accuracy in model 2 up to $\mu^2= 50\gev^2$ and
even down to $\mu^2= 2\gev^2$.  As shown in Fig.~\ref{fig:H-t}, this
is however not the case in model 1.  Here we find a clear departure
from an exponential behavior even when evolving from the starting
scale to $\mu^2= 6\gev^2$, whereas under backward evolution
$H^S(\xi,\xi,t)$ rapidly turns negative for some value of $t$.  We
notice that in model 1 the $x$ dependence of $H^S(x,0,t)$ at the
starting scale rapidly changes with $t$ due to the large value of
$\alpha'_v$.  This induces a corresponding change in the $x$
dependence of $H^S(x,\xi,t)$, which enters in the evolution equations.

To quantify the change of the $t$ dependence under evolution, we fit
the GPDs at given $\xi$ and $\mu$ to
\begin{equation}
  \label{fit-t}
H^i(\xi,\xi,t;\mu) = H^i(\xi,\xi,0;\mu)\,
  \exp\bigl[ t \bar{f}_i(\xi;\mu) \bigr]
\end{equation}
for $-t$ between $0$ and $0.5\gev^2$, where $i=g,S$.  Given the
behavior of the distributions under backward evolution, we restrict
these fits to $\mu^2\ge 4\gev^2$.  Whereas for $H^S$ in model 2 and
for $H^g$ in both models the form \eqref{fit-t} gives an excellent
description in the kinematical region of the fit, the corresponding
fit for $H^S$ in model~1 can only be approximate, as is seen in
Fig.~\ref{fig:H-t}.  This must be kept in mind when interpreting the
subsequent results, but despite this caveat the corresponding $t$
slope $\bar{f}_S$ gives a fair account of how $H^S(\xi,\xi,t)$ changes
with $\mu$.  The results of the fit are shown in
Fig.~\ref{fig:B-vs-xi} for the starting scale and for $\mu^2=
50\gev^2$.  We see that over a wide region of small $\xi$ the
dependence of $\bar{f}_i(\xi;\mu)$ on $\xi$ remains logarithmic after
evolution to higher scales.  For given $\mu$ we can hence perform a
fit
\begin{equation}
  \label{fit-xi}
\bar{f}_i(\xi;\mu) = \alpha'_i(\mu) \ln\frac{1}{\xi} + \bar{B}_i(\mu) \,.
\end{equation}
The results of such a fit in the range $3.2\times 10^{-5} <\xi<
3.2\times 10^{-4}$ are shown in Fig.~\ref{fig:apBxi}, where we plot
$\bar{f}_i(\xi;\mu)$ at the midpoint $\xi= 10^{-4}$ of the fit range,
as well as the effective shrinkage parameter $\alpha'_i(\mu)$.  In
model 2, $\bar{f}_i(\xi;\mu)$ and $\alpha'_i(\mu)$ are equal for the
gluon and the quark singlet to a good precision at the starting scale
by construction.  They change rather mildly under evolution to higher
scales, but a visible difference between gluon and singlet appears,
especially for $\alpha'_i$.  In model 1, we see that the slope and the
shrinkage parameter for the singlet evolve quite strongly and tend to
approach the corresponding values in the gluon distribution, which
increasingly dominates evolution with increasing $\mu$.  The
respective values for the gluon and the quark singlet are however
clearly different even at $\mu^2= 50\gev^2$.

\begin{figure}
\begin{center}
  \psfrag{k}[r][c]{\scriptsize $\mu^2= 50 \gev^2$}
  \psfrag{d}[r][c]{\scriptsize $6 \gev^2$}
  \psfrag{h}[r][c]{\scriptsize $4 \gev^2$}
  \psfrag{b}[r][c]{\scriptsize $2 \gev^2$}
  \psfrag{xx}[c][c]{\footnotesize $-t\; [\gev^2]$}
  \psfrag{yy}[c][t]{\footnotesize $H^{u-d}(\xi,\xi,t)$}
\includegraphics[width=\plotwidth]{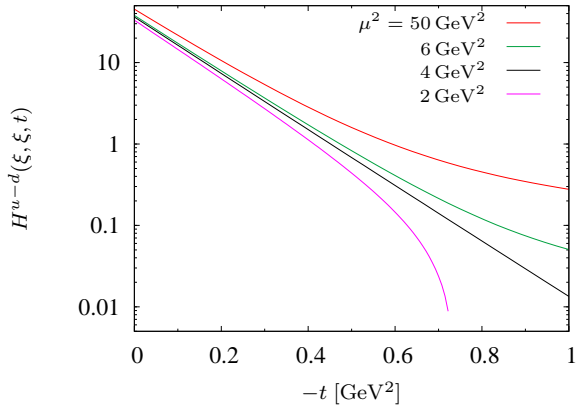}
\end{center}
\vspace{-1.5em}
\caption{\label{fig:Hud-t} $H^{u-d}(\xi,\xi,t)$ at $\xi= 3.2\times
  10^{-4}$ for different scales $\mu^2$.}
\end{figure}

We have also investigated the evolution behavior of the non-singlet
quantity $H^{u-d}(\xi,\xi,t)$.  Only the flavor difference enters for
the sea quark distributions \eqref{sea-quarks} in this case, and we
restrict our investigation to model~1.  The $t$ dependence changes in
a similar way as for the quark singlet in model 1, as becomes evident
from comparison of Fig.~\ref{fig:H-t} and Fig.~\ref{fig:Hud-t}.  In
particular, evolution to higher scales modifies the exponential
behavior of the initial condition.  A fit of the $t$ slope for $-t$
between $0$ and $0.5\gev^2$ must hence be taken with the same caveat
as above.  The $t$ slope $\bar{f}_{u-d}(\xi;\mu)$ fitted as in
\eqref{fit-t} shows an approximately logarithmic $\xi$ behavior in the
full range $2\gev^2 \le \mu^2 \le 50\gev^2$, so that we can again
perform a fit to the form \eqref{fit-xi} in the region $3.2\times
10^{-5} <\xi< 3.2\times 10^{-4}$.  Between $\mu^2= 2\gev^2$ and
$50\gev^2$, the resulting shrinkage parameter $\alpha'_{u-d}$
increases only by about 2\%, and the slope $\bar{f}_{u-d}(\xi;\mu)$
at $\xi= 10^{-4}$ decreases by about 10\%.  Compared with the quark
singlet sector in model 1, evolution effects in the non-singlet sector
are hence considerably weaker.


\section{Conclusions}
\label{sec:concl}

We have studied several aspects of the evolution behavior of GPDs at
small $\xi$.  To do this, we assumed a particular form of the GPDs at
a moderately low scale and numerically evolved this model ansatz to
higher $\mu$.
At $t=0$ we have taken initial conditions for which
$H(\xi,\xi,0)$ and the corresponding parton density $H(\xi,0,0)$
approximately obey power-laws with the same power.  Under evolution to
higher scales this power changes but remains the same for a GPD and
the associated PDF.  As a consequence, the skewness ratio $R(\xi,\mu)$
is only weakly $\xi$ dependent, to the extent that the effective power
changes with $\xi$.  The values of $R(\xi,\mu)$ for different initial
conditions approach each other with increasing $\mu$, and at high
scales they are well approximated by the Shuvaev formula
\eqref{shu-ratio}.  This convergence is however not very fast: with
rather different values of $R(\xi,\mu)$ at $\mu^2= m_c^2$ it only
becomes visible at $\mu^2$ of a few $10 \gev^2$ for the gluon and
quark singlet distributions, and at yet larger values in the
non-singlet sector.  We have not attempted to study how the situation
would change for initial conditions at much lower scale, considering
that in this case the leading-order approximation of the evolution
equations would no longer be suitable for drawing quantitative
conclusions.  We confirm the finding of \cite{Kirch:2005tt} that
evolution to higher scales generates a singular derivative
$(\partial/\partial x) H(x,\xi,t)$ at $x=\xi$ for quarks, but not for
gluons.

To study the change of $t$ dependence under evolution, we have chosen
initial conditions at $\mu_0= 2\gev$ such that $H(\xi,\xi,t) \sim
\exp[ t \bar{f}(\xi) ]$ and $\bar{f}(\xi) = \alpha' \ln(1/\xi) +
\bar{B}$ at small $\xi$ and $t$.  For distributions with a small
shrinkage parameter $\alpha'$, we find that to a good approximation
the $t$ dependence remains exponential under evolution to higher (and
to some extent also to lower) scales.  In contrast, a deviation from
an exponential $t$ behavior becomes visible after evolution rather
quickly for distributions with large $\alpha'$ at the starting scale,
so that a fit to an exponential form is only approximate in these
cases.  The fitted slopes $\bar{f}(\xi)$ of the evolved GPDs retain a
logarithmic $\xi$ dependence, so that one can also determine a
shrinkage parameter $\alpha'$ at different scales $\mu$.  We find that
the values of $\alpha'$ for the gluon and quark singlet distributions
remain close (but not equal) to each other under evolution in a model
where they coincide at $\mu_0$.  In an alternative model, where
$\alpha'$ for the quark singlet is much larger than for gluons at
$\mu_0$, evolution brings their values closer to each other, but clear
differences remain even at $\mu^2= 50\gev^2$.  An analogous behavior
is found for $\bar{f}(\xi)$ at given $\xi$ in both models.  We
therefore conclude that one may not take it for granted that the $t$
dependence of gluon and sea quark distributions is the same at
moderate scales.  In the flavor non-singlet sector, we find that
$\bar{f}(\xi)$ and $\alpha'$ remain quite stable under evolution of
the scale.


\section*{Acknowledgments} 

We thank R. Thorne for drawing our attention to the findings in
Ref.~\cite{Martin:1996as} and to their explanation within the double
logarithmic approximation \cite{DeRujula:1974rf}.  This work is
supported by the Helmholtz Association, contract number VH-NG-004.


\end{document}